\documentclass[preprint,tightenlines,eqsecnum,floats,aps,amsmath,amssymb,nofootinbib,prd,showpacs]{revtex4}

\usepackage{amssymb}
\usepackage{stmaryrd}
\usepackage{amsmath}
\usepackage{amsfonts}
\usepackage{mathrsfs}
\usepackage{CJK}
\usepackage{amsmath,amssymb,amsfonts}
\usepackage{graphicx}
\usepackage{subfigure}

\def\be{\begin{equation}}
\def\ee{\end{equation}}
\def\ba{\begin{eqnarray}}
\def\ea{\end{eqnarray}}
\def\nn{\nonumber}

\begin{document}

\title{Maximal efficiency of the collisional Penrose process with spinning particles in Kerr-Sen black hole}

\author{Yunlong Liu}
\affiliation{Department of Physics, South China University of Technology, Guangzhou 510641, China}

\author{ Xiangdong Zhang\footnote{Corresponding author. scxdzhang@scut.edu.cn}}
\affiliation{Department of Physics, South China University of Technology, Guangzhou 510641, China}

\date{\today}


\begin{abstract}
	We study the collision of two uncharged spinning particles around an extreme Kerr-Sen black hole and calculate the maximal efficiency of the energy extraction from the  Kerr-Sen black hole via super Penrose process. 	
	We consider the collision of two massive particles as well as collision of a massless  particle with a massive  particle. We calculate the maximum efficiency for all the cases, and found that the efficiency increases as the Kerr-Sen black hole's parameter($b=1-a$) decreases.
\end{abstract}
\maketitle
\section{INTRODUCTION}

Penrose process,  a mechanism to extract rotational energy from black hole, was first discovered by Penrose in 1969 with Kerr black hole\cite{Gravitational_collapse_Penrose}. The original version of Penrose process happened in the ergosphere, an object splits into two parts while it falls toward Kerr black hole. The one falls into the black hole with negative energy, while the other escapes  to infinity. The energy of escaped part is larger than the original one. Therefore,  rotational energy can be extracted from Kerr black hole. For such a process, Wald obtain the maximum efficiency $\eta_{max} = (output \enspace energy)/(input \enspace  energy) \approx 1.21$\cite{Energy_limits_Wald_1974}. After that, Piran et al. consider a different type of collision Penrose process which two particles collide inside the ergosphere, and found to be have similar energy contracting efficiency with the original Penrose process\cite{High_efficiency_Piran_Shaham}.

In 2009, Ba\~nados, Silk and West(BSW) proposed that a rotating black hole can act as accelerators for non-spin particles\cite{PhysRevLett.103.111102}. They show that the collision center-of-mass energy can be arbitrarily high for extremal Kerr black hole\cite{PhysRevLett.103.111102}. Inspired by this work, some authors suggest to construct Penrose process based on the BSW mechanism\cite{PhysRevLett.109.121101,PhysRevD.86.024027,PhysRevLett.113.261102}. These collision process are called as super Penrose process since it usually has far more higher energy contraction efficiency. For example, Schnittman obtain the maximal efficiency is about 13.92 when a massless and a massive particle collide near the horizon\cite{PhysRevLett.113.261102}. Along this line, the super Penrose process have been extended to various black holes\cite{PhysRevD.97.064024, PhysRevD.98.044006, PhysRevD.98.064027, Okabayashi:2019wjs}.

Recently, the BSW mechanism has been generalized to include the spinning particles\cite{Armaza2015Can,PhysRevD.94.124017,Zaslavskii16,PhysRevD.82.083004,DR18,PhysRevD.82.103005,PhysRevD.83.084041,PhysRevD.83.044013,PhysRevD.93.084025,PhysRevD.97.024003,PhysRevD.99.064022}. It has been shown \cite{A.Papapetrou.Spinning_test-particles,Dixon1970Dynamics1,Dixon1970Dynamics2,Wald1972Gravitational,Kerr1963Gravitational} that the trajectory
of a spinning test particle is no longer a geodesic and therefore is more close to the real particle. The corresponding super Penrose process also have been investigated in
many cases\cite{PhysRevD.97.064024, PhysRevD.98.044006, PhysRevD.98.064027, Okabayashi:2019wjs}. It worth to note that in\cite{Gao18}, the authors obtain some general result on the energy in the center of mass frame for BSW mechanism. However, this paper is devoted to the efficiency of Penrose process that was not studied in \cite{Gao18}.

On the other hand, the Kerr-Sen black hole is a rotating and charged solution of the low-energy effective field theory for heterotic string theory\cite{PhysRevLett.69.1006}. After it proposed, many aspects of Kerr-Sen solution has been investigated\cite{Gwak:2016gwj}. This black hole solution characterized by three parameters, which are mass $M$,  angular momentum $a$,  and charge $Q$($b=Q^2/2M$). It reduces to the Kerr black hole when the parameter $b=0$. As a grand unified theory, string theory is the most promising candidate of
unified all the interactions, to this sense, the expected rotating and charged black
hole solution would be the Kerr-Sen black hole rather than the
Kerr-Newman one. Therefore in this paper we investigate the issue of the maximal energy contraction efficiency of the super Penrose process for spin particles in Kerr-Sen background.  We provide a deep analysis of
the super Penrose process for spinning particles and investigate
the dependence of the maximal energy contraction efficiency with the Kerr-Sen black hole's parameter.

This paper is organized as follows: After an introduction, we discuss the equations of motion for spinning particles in Kerr-Sen black hole in Sec.II. While in Sec.III, we study the super Penrose collision of spinning particles in extreme Kerr-Sen  background. This section is divided into three cases and calculate the  the maximal efficiency with different parameters of extreme Kerr-Sen black hole.  The summary and conclusion was given in Sec.IV. Through out the paper, we adopt the geometrical unit $(\mathit{c}=\mathit{G}=1)$.

\section{BASIC EQUATION}

\subsection{Equations of motion of a spinning particle}
The equations of motion for spin particle in the curved spacetime can be described  by  Mathission-Papapetrou-Dixon(MPD) equations\cite{A.Papapetrou.Spinning_test-particles,Dixon1970Dynamics1,Dixon1970Dynamics2} 
\begin{eqnarray}
\frac{Dp^a}{D \tau}&=&-\frac{1}{2} R^{a}{}_{bcd} v^{b}S^{cd}\\
\frac{D S^{ab}}{D \tau}&=&p^a v ^b-p^b v^a
\end{eqnarray}
where
\begin{eqnarray}
v^a=(\frac{\partial}{ \partial\tau})^a
\end{eqnarray}
is the tangent vector of the center-of-mass world line, $\frac{D}{D\tau}$ is the covariant derivative along worldline, and
$ p^a=m u^a$ is the canonical 4-momentum of the spinning particles which satisfy
\begin{eqnarray}
p^a p_a=-m^2.
\end{eqnarray}
Moreover $S^{ab}$ is the particle's antisymmetric spin tensor, and its square turns out to be the spin of the particle as follows,
\begin{eqnarray}
S^{ab}S_{ab}=2S^2=2 m^2 s^2
\end{eqnarray}
where $s$ and $m$  are the spin and mass of the given particle respectively.
In the following, for the convenient of the calculation, we add a supplementary conditions between $S^{ab}$ and $P^a$ as follows
\begin{equation}
S^{ab}p_{a}=0
\end{equation}
Furthermore, we also normalize the affine paramenter $\tau$ through
\begin{equation}
u^a v_a=-1
\end{equation}
A detailed calculation shows a relation between $v^a$ and $u^a$ as
\begin{eqnarray}
v^a-u^a=\frac{S^{ab}R_{bcde}u^cS^{de}}{2(m^2+\frac{1}{4}R_{bcde}S^{bc}S^{de})}\label{vu}
\end{eqnarray}
The Eq. \eqref{vu} means that the 4-velocity and 4-momentum are not always parallel.
In addition, we can obtain the conserved quantities for spin particles with Killing vector fields $\xi_a$ as follows:
\begin{eqnarray}
Q_{\xi}=p^a \xi_a +\frac{1}{2}S^{ab} \nabla_a \xi_b
\end{eqnarray}

\subsection{Conserved quantities in the Kerr-Sen black hole}
Here we will consider the Kerr-Sen background and we can calculate the conserved quantities explicitly. In the Boyer-Lindquist coordinates $(t,r,\theta,\phi)$, the Kerr-Sen metric can be written as
\begin{eqnarray}
\label{dsmetric}
ds^2=-\frac{\Delta-a^2sin^2\theta}{\Sigma}dt^2+\frac{\Sigma}{ \Delta} dr^2+\Sigma d \theta^2+\frac{\Xi sin^2(\theta)}{\Sigma}d\phi^2-\frac{4Mrasin^2 \theta}{\Sigma}dtd\phi
\end{eqnarray}
where $\Sigma=r(r+2b)+a^2cos^2\theta$, $\Delta=r(r+2b)-2Mr+a^2$, $\Xi=(a^2+r(2b+r))^2-\Delta a^2sin^2\theta$ and $b=Q^2/2M$.

The nonvanishing components of the inverse metric $g^{\mu\nu}$ read
\begin{eqnarray}
&&g^{tt}={-\frac{\Xi  \Sigma }{a^2 \sin ^2(\theta ) (4 M^2 r^2-\Xi )+\Delta  \Xi }} \notag\\
&&g^{rr}={g^{\theta \theta} \Delta}=\frac{\Delta}{ \Sigma}  \notag\\
&&g^{\phi \phi}=\frac{\Sigma \csc^2(\theta ) (\Delta \csc ^2(\theta )-a^2)}{a^2 (4M^2 r^2-\Xi )+\Delta \Xi \csc^2(\theta )} \notag\\
&&g^{t \phi}=-\frac{2 a M r \Sigma }{a^2 \sin ^2(\theta ) \left(4 M^2 r^2-\Xi \right)+\Delta  \Xi }
\end{eqnarray}
In order to simplify the equation, we introduce a tetrad basis as
\begin{eqnarray}
&&e_a^{(0)}=\sqrt{\frac{\Delta}{\Sigma}}\left(dt_a-a sin ^2\theta d\phi_a \right)	\nn\\
&&e_a^{(1)}=\sqrt{\frac{\Delta}{\Sigma}}dr_a \nn\\
&&e_a^{(2)}=\sqrt{\Sigma} d\theta_a \nn\\
&&e_a^{(3)}=\frac{\sin \theta }{\sqrt{\Sigma }}\left(-adt_a + (r^2+2br+a^2)d\phi_a\right) \label{tetrad}
\end{eqnarray}
There exist two Killing vectors in the Kerr-Sen geometry:
\begin{eqnarray}
\xi^a=\left(\frac{\partial}{\partial t}\right)^a;    \quad  \phi^a=\left(\frac{\partial}{\partial \phi}\right)^a
\end{eqnarray}
Then the conserved quantities in Kerr-Sen background associated to the above two Killing vectors can be written as
\begin{small}
\begin{eqnarray}
\label{Energyoftheparticle}
E&=&-Q_{\xi^t}= \sqrt{\frac{\Delta }{\Sigma }} p^{(0)}+\frac{ a \sin \theta}{\sqrt{\Sigma }}p^{(3)}-\frac{M \left(r^2-a^2 \cos ^2\theta \right)}{\Sigma ^2}S^{(01)}+\frac{2 a M r \cos \theta }{\Sigma ^2}S^{(23)}\\
\label{Angularmomentumoftheparticle}
J&=&Q_{\xi ^{\phi }}= a \sin^2 \theta \sqrt{\frac{\Delta }{\Sigma }} p^{(0)}+\frac{ \left(a^2+r (2 b+r)  \right){\sin\theta}}{\sqrt{\Sigma }}p^{(3)}-\frac{ a \sin ^2 \theta }{\Sigma ^2}\left(b (2 M r+\Sigma )+2 M r^2 +(r-M) \Sigma \right)S^{(01)}  \notag\\
&&-\frac{a \sqrt{\Delta }  cos\theta \sin \theta }{\Sigma }S^{(02)}+\frac{(b+r)\sqrt{\Delta } sin\theta  }{\Sigma }S^{13} + \frac{ cos\theta }{\Sigma ^2}\left(\left(a^2+r (2 b+r)\right)^2-a^2 \Delta \sin ^2\theta \right)S^{(23)}
\end{eqnarray}
\end{small}
where $E$  and $J$ are the energy and angular momentum of the particle respectively.
\subsection{Equations of motion on the equatorial plane}
When the particle's spin is aligned with the spin of the black hole, the spin $s^{(a)}$ can be show as follow:
\begin{eqnarray}
s^{(a)}=-\frac{1}{2m}\varepsilon^{(a)}{}_{(b)(c)(d)}u^{(b)}S^{(c)(d)}
\end{eqnarray}
equivalently
\begin{eqnarray}
S^{(a)(b)} = m \varepsilon^{(a)(b)}{}_{(c)(d)}u^{(c)}s^{(d)}
\end{eqnarray}
where $\varepsilon_{(a)(b)(c)(d)}$ is the completely antisymmetric tensorn, with component  $\varepsilon_{(0)(1)(2)(3)}=1$. Furthermore, we consider that the particle
was confined  in the  equatorial plane($\theta=\pi /2$) \cite{PhysRevD.58.064005}.
The non-zero components of spin tensor read
\begin{eqnarray}
\label{ValueofspinS}
S^{(0)(1)}=-sp^{(3)}, \quad
S^{(0)(3)}= sp^{(1)}, \quad
S^{(1)(3)}= sp^{(0)}
\end{eqnarray}
Combining Eqs \eqref{Energyoftheparticle}, \eqref{Angularmomentumoftheparticle}, and \eqref{ValueofspinS}, the  equations of momentum can be written as
\begin{small}
	\begin{eqnarray}
	\label{MMMformulamotionequationp0}
	p^{(0)}&=&\frac{\Sigma}{\mathcal{D}_1} \Big( -J \left(a \Sigma \sqrt{\Sigma }+M r^2 s\right) \notag \\
	&&+E \left(\Sigma \sqrt{\Sigma } \left(a^2+r (2 b+r)\right)+a s \Sigma (b-M+r)+2 a M r s (b+r)\right)\Big)\\
	\label{MMMformulamotionequationp3}
	p^{(3)}&=&\frac{\Sigma ^2\sqrt{\Delta }}{\mathcal{D}_1} \left( J \sqrt{ \Sigma }- E \left( s (b+r)+a \sqrt{ \Sigma }\right) \right)
	\end{eqnarray}
\end{small}
where
\begin{eqnarray}
\label{MMMformulamotionequationpD}
\mathcal{D}_1  = a M s \sqrt{\Delta  \Sigma } \left(2 b r+r^2-\Sigma \right)+r \sqrt{\Delta } \left(\Sigma ^2 (2 b+r)-M r s^2 (b+r)\right)
\end{eqnarray}
there are a normalization condition of the 4-momentum as \cite{PhysRevLett.113.261102}
\begin{eqnarray}
\label{normalizationcondition}
p^{(a)}p_{(a)}=k
\end{eqnarray}
where  $k=-m^2$ for the massive and  $k=0$ for massless particles. As for massive particles, we defined  a specific 4-momentum $u^{(a)}$, by $u^{(a)}=p^{(a)}/m$.
Hence, with Eq.\eqref{normalizationcondition} in hand, for the massive particles, we have
\begin{eqnarray}
\label{expressionofu1}
u^{(1)}=\sigma \sqrt{(u^{(0)})^2-(u^{(3)})^2-1}
\end{eqnarray}
Here $\sigma=\pm1$ denote the outgoing and ingoing
motions respectively. Moreover, combining Eqs.\eqref{vu}, \eqref{ValueofspinS} and \eqref{dsmetric},  the expressions of the 4-velocity read
\begin{eqnarray}
v^{(0)}&=&\frac{1}{\mathcal{D}_2}  \left( u^{(0)} ( \mathcal{F}_1 + f_{v1} )+ u^{(3)} \mathcal{F}_2  \right)\\
v^{(1)}&=&\frac{1}{\mathcal{D}_2}  u^{(1)} \mathcal{F}_1 \label{v1}\\
v^{(3)}&=&\frac{1}{\mathcal{D}_2}  \left( u^{(3)}  (\mathcal{F}_1+f_{v2})- u^{(0)} \mathcal{F}_2 \right)
\end{eqnarray}
where
\begin{small}
	\begin{eqnarray}
	f_{v1}&\!=\!&b^2 s^2 \Delta;
	 \quad
	 f_{v2}=s^2 \big(M r^2 (b+3 r)-a^2 b^2\big);\\
	 \label{mathcalF_1}
	\mathcal{F}_1&\!=\!&r^2 \big(r (2 b + r)^3 - M (b + r) s^2 \big);
	 \quad
	\mathcal{F}_2=a b^2 \sqrt{\Delta } s^2;\\
	\mathcal{D}_2&\!=\!&
	r^6 +r^2 \Bigg(-b M s^2 \left((u^{(3)})^2+1\right)+6 b r^3-M r s^2 \left(3 (u^{(3)})^2 +1\right)\Bigg)+2 b^3 \left(4 r^3+r s^2 (u^{(0)})^2\right)+ \notag\\
	&&
	b^2 \Bigg(a s^2 \left(a \left((u^{(0)})^2+(u^{(3)})^2\right)+2 \sqrt{\Delta } u^{(0)} u^{(3)}\right)-2 M r s^2 (u^{(0)})^2+12 r^4+r^2 s^2  (u^{(0)})^2 \Bigg)
	\end{eqnarray}
\end{small}
By employing the tetrad basis \eqref{tetrad}, 4-velocity can be  rewritten as
\begin{eqnarray}
\frac{dt}{d \tau}&=&\frac{\left(a^2+\Sigma \right)v^{(0)} +a \sqrt{\Delta } v^{(3)} }{\sqrt{\Delta \Sigma }}\\
\frac{dr}{d \tau}&=&\frac{\sqrt{\Delta } v^{(1)} }{\sqrt{\Sigma}}\label{drdtau}\\
\frac{d \varphi}{d \tau}&=&\frac{a v^{(0)} +\sqrt{\Delta } v^{(3)}}{\sqrt{\Delta  \Sigma}}
\end{eqnarray}
By plugging \eqref{v1} to Eq.\eqref{drdtau},  the  radial equation of motion for spin particle gives rise to
\begin{eqnarray}
\frac{dr}{d \tau}=\frac{\mathcal{F}_2 \sqrt{\Delta } }{\mathcal{D}_2 \sqrt{\Sigma}} u^{(1)}
\label{radial_equation_of_motion}
\end{eqnarray}
In order to facilitate the numerical calculation and without loss generality, we simply set the variables to the dimensionless variables as
\begin{eqnarray}
    \tilde {E} = \frac{E}{m}
, \ \tilde {J} = \frac{J}{m M}
, \ \tilde {s} = \frac{s}{M}
, \ \tilde {t} = \frac{t}{M}
, \ \tilde {r} = \frac{r}{M}
, \ \tilde {a} = \frac{a}{M}
, \ \tilde {Q} = \frac{Q}{M}
, \ \tilde {\tau} = \frac{\tau}{M}
\end{eqnarray}
This is equivalent to discuss the energy and other quantity with unity mass. In the following, we omit the $\ \widetilde{}$ for simplicity. For example, $E$ in the following text actually
means $\tilde {E} $.
\subsection{Constraints on the orbits}
In this part, we devoted to find the admissible trajectory of the spin particle which can apporoch to the horizon $r_H=\sqrt{(1-b)^2-a^2}-b+1$, this means that  the equation \eqref{radial_equation_of_motion} must be have real solutions. Combining this fact with the Eq. \eqref{expressionofu1}  gives us
\begin{eqnarray}
(u^{(0)})^2-(u^{(3)})^2-1 \ge 0
\end{eqnarray}
when $r \ge r_H$. By plugging Eq.\eqref{MMMformulamotionequationp0} and \eqref{MMMformulamotionequationp3} into the above equation, we get a constraint on the orbits
\begin{eqnarray}
E^2 \bigg(\Sigma ^2 \left(\Sigma ^{3/2} \left(a^2-a B_r+r (2 b+r)\right)+ r s (2 a (b+r)-B_r r)+a s \Sigma  (b-1+r)\right)^2      \notag\\
  -\Sigma ^4 \left((a-B_r) \sqrt{\Delta  \Sigma }+\sqrt{\Delta } s (b+r)\right)^2 \bigg) \geq \mathcal{D}_1^2\label{orbitcondition1}
\end{eqnarray}
where $B_r=J/E$ and $\mathcal{D}_1$ is given at Eq. \eqref{MMMformulamotionequationpD}.
For the orbits which can reach the horizon, note that $\Delta$ and $\mathcal{D}_1$ vanished at the horizon($r=r_H$),  Eq.\eqref{orbitcondition1} gives a critical value of $B_r$
\begin{eqnarray}
\label{section1B_cr}
\!B_{cr}\!=\!\frac{-a^2 r_H (2 b+r_H)^2-a s \sqrt{r_H (2 b+r_H)} ((b+r_H) (2 b+r_H)+r_H)-r_H^2 (2 b+r_H)^3}{-a r_H (2 b+r_H)^2-s r_H \sqrt{r_H (2 b+r_H)}}
\end{eqnarray}
Hence the condition that the orbit can reach the horizon equal to $B_r \le B_{cr}$.
On the other hand, we know that for a massive particle, the 4-velocity along the admissible trajectory must be timelike as
\begin{eqnarray}
v_{(a)} v^{(a)}=-(v^{(0)})^2+(v^{(1)})^2+(v^{(3)})^2<0
\end{eqnarray}
Along the same line of \cite{PhysRevD.98.064027}, the above timelike condition is equivalent to the following constraint
\begin{eqnarray}\label{section1ConditionofE}
\mathcal{U} E^2<{\mathcal{C}_a}^2 r^2 {\mathcal{F}_1}^2
\end{eqnarray}
where ${\mathcal{C}_a}=r (2 b+r)^3-s^2 (b+r)$,  $\mathcal{F}_1$ is at Eq.\eqref{mathcalF_1} , and the detailed expression for $\mathcal{U}$ can be found in APPENDIX.

\begin{figure}[!htb]
	\centering
	\subfigure[{}]{
		\label{fig:conditionofthespin} 
		\includegraphics[width=0.48\textwidth]{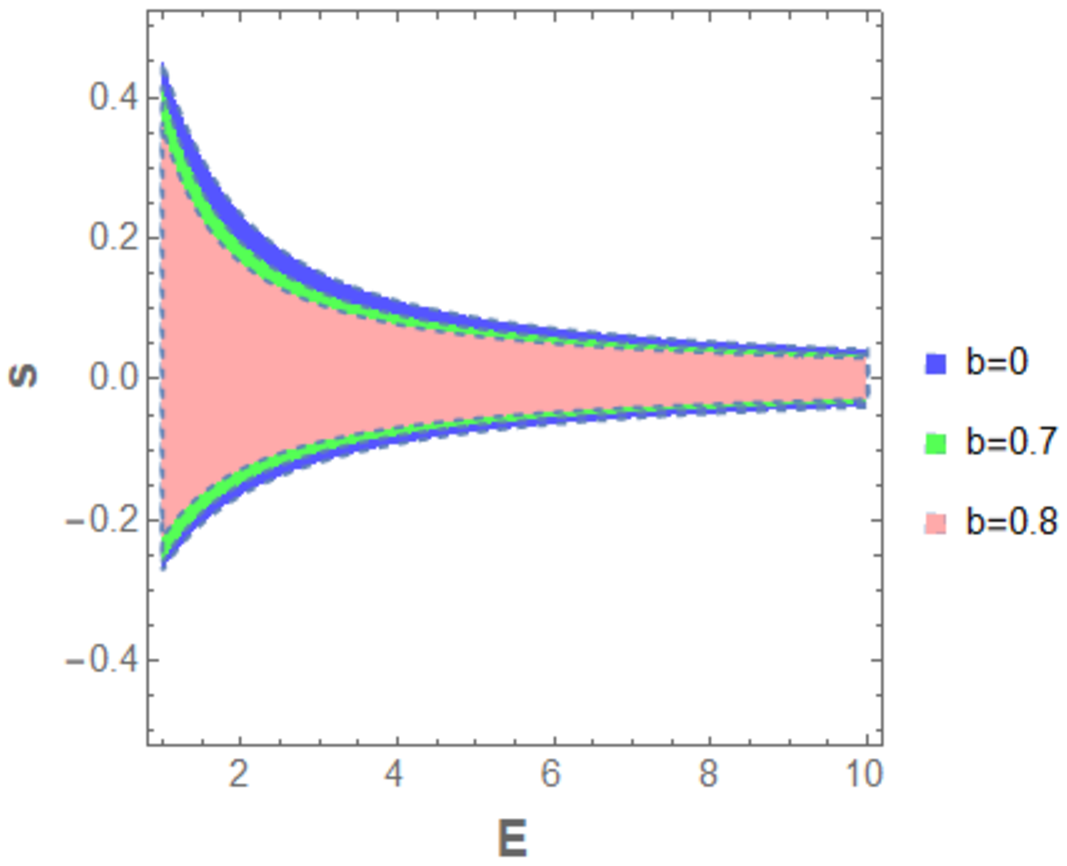}
	}
	\subfigure[{}]{
		\label{fig:conditionofxi} 
		\includegraphics[width=0.48\textwidth]{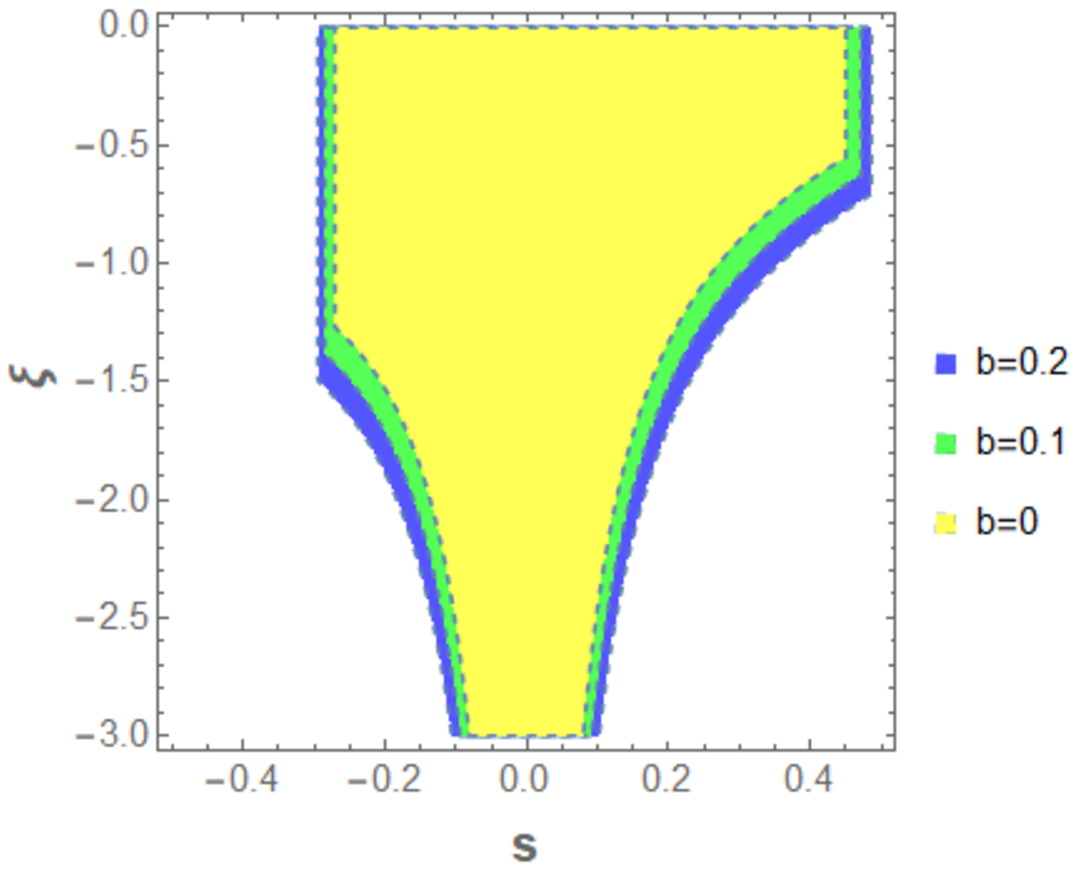}
	}
	\caption{	(a)The condition of the spin $s$ and the energy $E$ that the particles can reach the  event horizon for different value of $b$.
		(b) The condition of the spin $s$ and the $\xi$ that the particles can fall into the horizon for different value of $b$.}
\end{figure}

\begin{figure}[!htb]
	\includegraphics [width=0.4\textwidth]{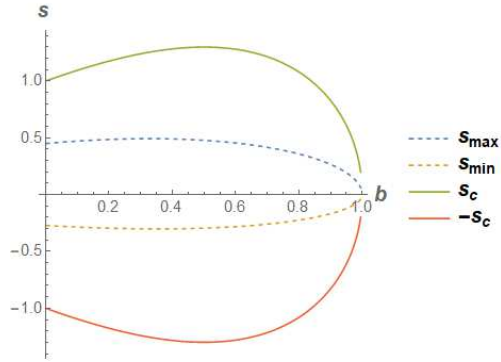}
	\caption{ The maximal and minimum value of spin $s$ of the particles in the extreme case for different values of $b$ as well as its comparison with $s_c$.}
	\label{conditionofsFordiffrerentb}
\end{figure}

Since we consider the maximal energy contraction efficiency from black hole,  in the following we only focus on the extreme Kerr-Sen black hole($b=1-a$). In this situation, if one of the collision particle possess the critical angular momentum, it is easy to see
that $B_{cr}=2$ from Eq.\eqref{section1B_cr}.
Then from Eq. \eqref{section1ConditionofE}, we have
\begin{small}
\begin{eqnarray}\label{section1extremeConditionofE}
E^2<\frac{\left((b-1) (b+1)^3+s^2\right)^4}{(b-1) (b+1)^4 (b+3) s^2 \left((b+1) \sqrt{1-b^2}-s\right)^2 \left((b (b+2)-1) s^2+2 (b-1) (b+1)^3\right)}
\end{eqnarray}
\end{small}
which gives us a constraint on energy $E$ for different values of spin $s$ and $b$ and is showed in the Fig. \ref{fig:conditionofthespin}. The figure shows that when $b$ increase, the admissible range of spin $s$ shrinks for a given energy $E$.
If the particle falling from infinity, that is $E\ge 1$, combining this fact with Eq. \eqref{section1extremeConditionofE},  the spin $s$ will be restricted to $s_{min}<s<s_{max}$  for a
given value of $b$. For example, when $b=0.1$, we can obtain $s_{min}\approx -0.285$ and $s_{max}\approx 0.471$. More information of $s_{min}$ and $s_{max}$ for different value of $b$ can be found in Fig. \ref{conditionofsFordiffrerentb}.
Moreover, it worth to note that, the authors of Ref\cite{Gao18} point out that when the particle process critical spin $s=-s_c=-a^2\left(\frac{2}{a}-1\right)^{\frac32}$, the timelike condition
is violated. We show in Fig. \ref{conditionofsFordiffrerentb}, our admissible spin $s$ corresponding to the maximum of efficiency is always bigger than critical value ($s>-s_c$),
and therefore the timelike condition is satisfied in our case.

If the particle's angular momentum is deviate from critical value,  we set $B_r=2(1+\xi)$ with $\xi$ being a negative number. From Eq.\eqref{section1ConditionofE}, the energy $E$ now is a function of the $s$, $b$ and $\xi$ and is showed as the Fig. \ref{fig:conditionofxi}. This figure shows that the allowed range of $\xi$ increase when $b$ increases.

We assume the particles are freely falling from infinity. If $B_r>B_{rc}$, such a particle falling from
infinity will find a turning point away from horizon, and then bounce back to
infinity. So if $B_r=B_{rc}+\delta (\delta \rightarrow 0^+)$, the turning point of the particle can very close to the horizon. Then, these particles will moving outward. Therefore this situation should  also need to be taken into account.

\section{COLLISION OF SPINNING PARTICLES}
In this section, we  consider the collision of two spin particles that are freely falling from infinity, and find the formula of the efficiency of the energy extraction from the extreme  Kerr-Sen black hole.

We denotes that the 4-momentum of particle 1 and particle 2 are $p_1{}^\mu$ and $p_2{}^\mu$. Our picture is the following: The particles collide outside the horizon. After collision, particle 3, whose 4-momenta is $p_3{}^\mu$, will move to infinity, while the particle 4 with $p_4{}^\mu$ falls into the Kerr-Sen black hole. We assume that the sum of  initial spins and 4-momenta are conserved throughout the collision process. That is,
\begin{eqnarray}
\label{MMMformulaspinsconservation}
S_1^{\mu \nu}+S_2^{\mu \nu}&=&S_3^{\mu \nu}+S_4^{\mu \nu}\\
\label{MMMformula4-momentasconservation}
p_1^\mu+p_2^\mu&=&p_3^\mu+p_4^\mu
\end{eqnarray}
Since the Kerr-Sen spacetime exists two Killing vectors, contracting these two Killing vector with the above equation gives the conservation of the energy $E$ and angular momentum $J$ as follows
\begin{eqnarray}
E_1+E_2&=&E_3+E_4 \label{conservationE}\\
J_1+J_2&=&J_3+J_4  \label{conservationJ}
\end{eqnarray}
From the Eq. \eqref{MMMformulaspinsconservation} and \eqref{MMMformula4-momentasconservation}, we can also obtain the conservation of particle's spin and the radial components of 4-momentum throughout the collision
\begin{eqnarray}
m_1 s_1+m_2 s_2&=&m_3 s_3+m_4 s_4 \label{conservations}\\
p_1^{(1)}+p_2^{(1)}&=&p_3^{(1)}+p_4^{(1)}\label{conservationp}
\end{eqnarray}

Now we assume that particle 1 and particle 2 collide near the horizon of extreme Kerr-Sen black hole, the radial position of collision point $r_c$ is very close to extreme Kerr-Sen black hole's horizon $r_H(r_H=a=1-b)$, so that we can assume $(r_c=a/(1-\epsilon))$ with $\epsilon \rightarrow 0^+$. Then, we expand the particles' radial 4-momentum in terms of $\epsilon$ as follows:
\begin{eqnarray}
p^{(1)} =\sigma \frac{  (b+1) \sqrt{2 \sqrt{1-b^2} (b+1) s+(1-b) (b+1)^3+s^2} \left| J-2 E\right| }{\epsilon  \left((1-b) (b+1)^3-s^2\right)}+O(\epsilon^0)\label{p1}
\end{eqnarray}

In the following analysis, without loss of generality, along the same line of \cite{PhysRevD.98.064027}, we doing calculation in case that particle 1 is critical $(J_1=2 E_1)$, while particle 3 is near-critical$(J_3=2 E_3+O(\epsilon))$ and particle 2 is non-critical$(J_2<2 E_2)$ \cite{PhysRevD.98.064027}.

Then the total angular momentum of the particle can relate to the energy as follows:
\begin{eqnarray}
\label{MMMformulaBcrparticle1}
J_1&=&2 E_1\\
\label{MMMformulaBcrparticle3}
J_3&=&2 E_3 \left(1 + \alpha_3 \epsilon +\beta_3 \epsilon^2 +... \right)
\end{eqnarray}
where $\alpha_3$ and $\beta_3$ are expansion parameters of $O\left(\epsilon ^0\right)$.

For particle 2, since it is non-critical, we assume that:
\begin{eqnarray}
\label{MMMformulaBcrparticle2}
J_2&=&2 E_2 \left(1 + \xi \right)
\end{eqnarray}
where $\xi<0$ and $\xi=O(\epsilon^0)$

From the conservation law \eqref{conservationE} and \eqref{conservationJ}, we get the following equations
\begin{eqnarray}
E_4=E_1+E_2-E_3;\quad J_4=J_1+J_2-J_3;
\end{eqnarray}
which give us:
\begin{eqnarray}
\label{MMMformulaBcrparticle4}
J_4=2 E_4\left( 1+ \frac{E_2}{E_4} \xi +...  \right)
\end{eqnarray}

Since we consider the collision of the particle 1 and particle 2,  the particle 2 must be ingoing ($\sigma_2=-1$) because the particle 2 is noncritical\cite{PhysRevD.98.064027}. Combine Eq.\eqref{p1} with the conservation of 4-momentum \eqref{conservationp}, we can get the equation as follows:
\begin{small}
\begin{eqnarray} \label{section1p1-1orderfunction}
\left| J_2-2 E_2\right|  \left(\frac{\sigma_4 \sqrt{\mathcal{C}_a(s_4,b)+\mathcal{C}_b(s_4,b)}}{\mathcal{C}_a(s_4,b)}-\frac{\sigma_2 \sqrt{\mathcal{C}_a(s_2,b)+\mathcal{C}_b(s_2,b)}}{\mathcal{C}_a(s_2,b)}\right)=O\left(\epsilon ^1\right)\label{epsilonminus1}
\end{eqnarray}
\end{small}
where $\mathcal{C}_a(s,b)=(1-b) (1+b)^3-s^2$ is the critical case($r_H=1-b$) of the $\mathcal{C}_a$ in Eq. \eqref{section1ConditionofE} and $\mathcal{C}_b(s,b) = 2 (b+1) \sqrt{1-b^2} s+2 s^2$. From Eq.\eqref{epsilonminus1}, we find that $\sigma_4=\sigma_2$ and $s_4=s_2$. Then Eq.\eqref{conservations} further forces us to impose $s_3=s_1$.

In the following section, we will consider three different types of collision. The first case is the collision of two massive particles(MMM). The second type is the collision of one massless particle with another massive particle, which is called as compton scattering(PMP)\cite{PhysRevD.98.064027} and third type is the inverse compton scattering(MPM) \cite{PhysRevD.98.064027}, which is the inverse process of type two case.

Now, we come to calculate $E_2$ and $E_3$ for the cases [A](MMM), [B](MPM), and [C](PMP).

\subsection{Maximal Efficiency in Case [A] MMM}

For the case[A], to simplify the discussion, we just assume that the mass of collision particles are all equal to $m$, i.e. $m_1=m_2=m_3=m_4=m$.
With this in hand, the equations of conservation laws \eqref{conservationE}-\eqref{conservationp} can be simplified as
\begin{eqnarray}
E_1+E_2&=&E_3+E_4\\
J_1+J_2&=&J_3+J_4\\
s_1+s_2&=&s_3+s_4\\
u_1^{(1)}+u_2^{(1)}&=&u_3^{(1)}+u_4^{(1)}\label{conservationmmmu1}
\end{eqnarray}

The radial component of the 4-momentum of massive particle can be calculated from the Eqs. \eqref{MMMformulamotionequationp0} - \eqref{MMMformulamotionequationpD}, and \eqref{expressionofu1}. With the help of Eqs. \eqref{MMMformulaBcrparticle1}-\eqref{MMMformulaBcrparticle2}, and \eqref{MMMformulaBcrparticle4}, we expand the particles' radial 4-momentum in terms of $\epsilon$ as follows:
\begin{eqnarray}
\label{MMMformuluofaparticle1}
u_1^{(1)}=f_{11} \epsilon^{-1}+f_{12} \epsilon^{0}+f_{13} \epsilon^{1}+...\\
\label{MMMformuluofaparticle2}
u_2^{(1)}=f_{21} \epsilon^{-1}+f_{22} \epsilon^{0}+f_{23} \epsilon^{1}+...\\
\label{MMMformuluofaparticle3}
u_3^{(1)}=f_{31} \epsilon^{-1}+f_{32} \epsilon^{0}+f_{33} \epsilon^{1}+...\\
\label{MMMformuluofaparticle4}
u_4^{(1)}=f_{41} \epsilon^{-1}+f_{42} \epsilon^{0}+f_{43} \epsilon^{1}+...
\end{eqnarray}

From Eqs. \eqref{MMMformuluofaparticle1}-\eqref{MMMformuluofaparticle4}, we can easily obtain corresponding equations for different  order of $\epsilon$. Note that the leading order equation of $\epsilon^{-1}$ has already been discussed in the Eqs. \eqref{section1p1-1orderfunction} and we found some constraints have to be satisfied under Eq. \eqref{epsilonminus1}.
So we further discuss the Eq.\eqref{conservationmmmu1} from the next leading order of $\epsilon^{0}$ and $\epsilon^{1}$ as follows
\begin{eqnarray}
\label{MMMformula0order}
f_{12} +f_{22}=f_{32}+f_{42}\\
\label{MMMformula1order}
f_{13} +f_{23}=f_{33}+f_{43}
\end{eqnarray}
where
\begin{small}
\begin{eqnarray}
f_{12}&=&\frac{\sigma _1 k_1(E_1,s_1,b,0)}{f_1(s_1,b)}\\
f_{13}&=&-\frac{E_1^2 \sigma_1 k_3(s_1,b)}{f_1(s_1,b)^2 k_1(E_1,s_1,b,0)}\\
f_{22}&=&\frac{E_2 k_2(s_2,b,\xi )}{f_1(s_2,b)^2}\\
f_{23}&=&\frac{(1-b) f_1(s_2,b)^4-E_2^2 h_{41}(s_2,b,\xi )}{4 \sqrt{1-b^2} E_2 \xi  f_1(s_2,b)^3 f_2(s_2,b)}\\
f_{32}&=&\frac{\sigma_ 3 k_1(E_3,s_1,b,\alpha_3)}{f_1(s_1,b)}\\
f_{33}&=&-\frac{E_3^2 \sigma_3 (\beta_3 h_{61}(s_1,b,\alpha_3)+h_{62}(s_1,b,\alpha_3)+k_3(s_1,b))}{f_1(s_1,b)^2 k_1(E_3,s_1,b,\alpha_3)}\\
f_{42}&=&\frac{E_2 k_2(s_2,b,\xi )-f_1(s_2,b) (E_1 h_{71}(s_2,b)+E_3  h_{72}(s_2,b,\alpha_3))}{f_1(s_2,b)^2}\\
f_{43}&=&-\frac{1}{4 (b+1) \sqrt{1-b^2} E_2 \xi  f_1(s_2,b)^3 f_2(s_2,b)} \bigg( 2 (b+1) E_2 f_1(s_2,b) (4 E_3 \xi  h_{85}(s_2,b,\alpha_3,\beta_3) \notag \\
&&- (E_1-E_3) ((b-1) f_1(s_2,b) h_{81}(s_2,b)+2 \xi h_{83}(s_2,b)))  \notag\\
&&+ (b-1) (b+1) f_1(s_2,b)^2 \left(f_1(s_2,b)^2-(E_1-E_3)^2 h_{81}(s_2,b)\right)  \notag\\
&&+ E_2^2 \left(-4 (b+1) \xi f_1(s_2,b) h_{83}(s_2,b)+h_{82}(s_2,b)+\xi^2 h_{84}(s_2,b)\right) \bigg)
\end{eqnarray}
\end{small}
where $k_1(E_1,s_1,b,0)$, $k_2(s_2,b,\xi )$, $h_1(s_1,b)$, $h_2(s_2,b)$ $h_{71}(s_1,b)$, $g_3(b,s_2,\xi )$ and so on are the functions of different parameters and we will show them in the appendix.

From the Eq.\eqref{MMMformula0order}, with the detailed expressions given by the above, we obtain the equation of $E_3$ as follow
\begin{eqnarray}
	\mathcal{A}_1 E_3^2-2 \mathcal{B}_1 E_3+\mathcal{C}_1=0
	\label{QuadraticfunctionOfE3}
\end{eqnarray}
where
\begin{small}
\begin{eqnarray}
\label{MMMformulaABC}
\mathcal{A}_1&=&\frac{f_1(s_1,b)^2 h_{72}(s_2,b,\alpha_3 )^2} {f_1(s_2,b)^2}-k_{12}(s_1,b,\alpha_3) \notag\\
\mathcal{B}_1&=&-\frac{h_{72}(s_2,b,\alpha_3)}{f_1(s_2,b)^2} \left(\sigma_1 f_1(s_1,b) f_1(s_2,b)
 k_1(E_1, s_1, b, 0) +E_1 f_1(s_1,b)^2 h_{71}(s_2,b)\right) \notag\\
\mathcal{C}_1&=&\frac{2 E_1 \sigma_1 f_1(s_1,b) h_{71}(s_2,b) k_1(E_1, s_1, b, 0)} {f_1(s_2,b)}
+E_1^2 \Big( k_{12}(s_1,b,0) + \frac{ h_{71}(s_2,b)^2 f_1(s_1,b)^2}{f_1(s_2,b)^2} \Big)
\end{eqnarray}
\end{small}

From Eqs.\eqref{QuadraticfunctionOfE3} and \eqref{MMMformulaABC}, we find that $\sigma_3$ is decoupled. So the sign of $\sigma_3$ will not affect the value of $E_3$. Since the  quadratic equation of $E_3$ \eqref{QuadraticfunctionOfE3} has two solutions. The larger solution of $E_3=E_{3,+}$ gives larger efficiency because the efficiency depends on the value of $E_3$ that will became explicit in following parts. Therefore, it is sufficient to consider the case of $\sigma_3=-1$ with the larger solution of $E_3=E_{3,+}$.
In conclusion, we can get the expression of $E_3$ and $E_2$ from the Eqs. \eqref{MMMformula0order} and \eqref{MMMformula1order}.
\begin{eqnarray}\label{MMMformulaE3}
E_{3,+}&=&\frac{\sqrt{\mathcal{B}_1^2-\mathcal{A}_1 \mathcal{C}_1}+\mathcal{B}_1}{\mathcal{A}_1}
\end{eqnarray}
and
\begin{eqnarray}\label{MMMformulaE2}
E_2&=&\frac{(b-1) (E_1-E_3)^2 }{\mathcal{P}_1}
\end{eqnarray}
where
\begin{small}
\begin{eqnarray}\label{MMMformulaP}
\mathcal{P}_1&=&\frac{2}{h_1(s_1,b)^2 h_{41}(s_2,b)} \bigg(-4 E_3 \xi  h_{85}(s_2,b,\alpha_3,\beta_3)+2 \sqrt{1-b^2} \xi  \big(f_{13}+f_{33}\big) f_1(s_2,b)^2 f_2(s_2,b) \notag\\
&&+ (b-1) (E_1-E_3) f_1(s_2,b) h_{81}(s_2,b)+2 \xi  (E_1-E_3) h_{83}(s_2,b) \bigg).
\end{eqnarray}
\end{small}

With all those ingredients, the efficiency can be calculated through the following expression:
\begin{eqnarray}
\eta=\frac{E_3}{E_1+E_2}
\end{eqnarray}

\subsubsection{Efficiency}
With the detailed expressions of $E_3$ and $E_2$ above. We have three different types of parameters involved in the calculation of the efficiency $\eta$. The first type is the charge of extreme Kerr-Sen black hole($b=1-a$). Second type is particles spins($s_1$ and $s_2$), the third type is orbit parameters of the particles such as ($\alpha_3$, $\beta_3$ and $\xi$) and direction of the particles' motion ($\sigma_1$, $\sigma_2$, $\sigma_3$ and $\sigma_4$).

Note that we already fix the value of $\sigma_2,\sigma_3,\sigma_4$ as $\sigma_2=\sigma_4=-1$ and $\sigma_3=-1$ in the last section. So the only remaining parameter for the direction of the particles' motion is $\sigma_1$. However, a good  efficiency can't be found for $\sigma_1=-1$\cite{PhysRevD.98.064027}, so  we set that  $\sigma_1=1$.

Then, for a given value of $E_1$, the maximal efficiency $\eta_{max}$ would be reached with the minimum value of $E_2$ and the maximal value of $E_3$. Without loss of generality, we just normalize the ingoing energy $E_1$ as $E_1=1$.

\begin{figure}[!htb]
	\centering
	\subfigure[{}]{
		\includegraphics[width=0.31\textwidth]{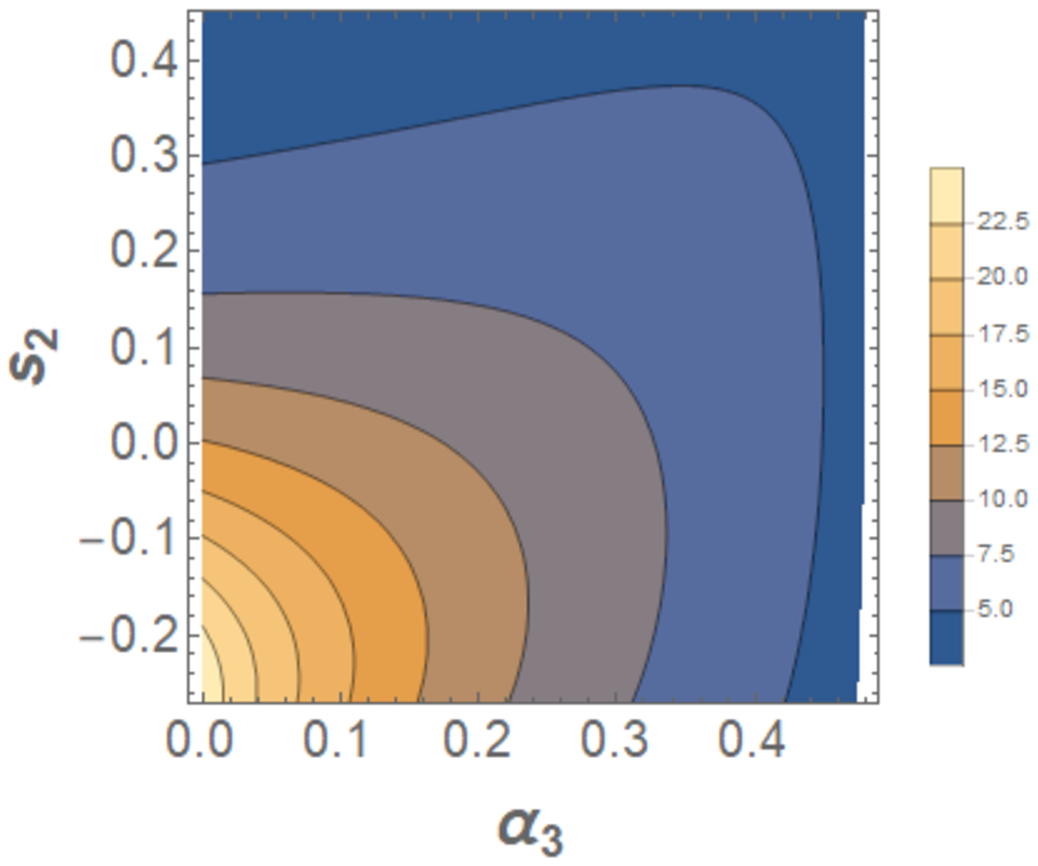}
	}
	\subfigure[{}]{
		\includegraphics[width=0.31\textwidth]{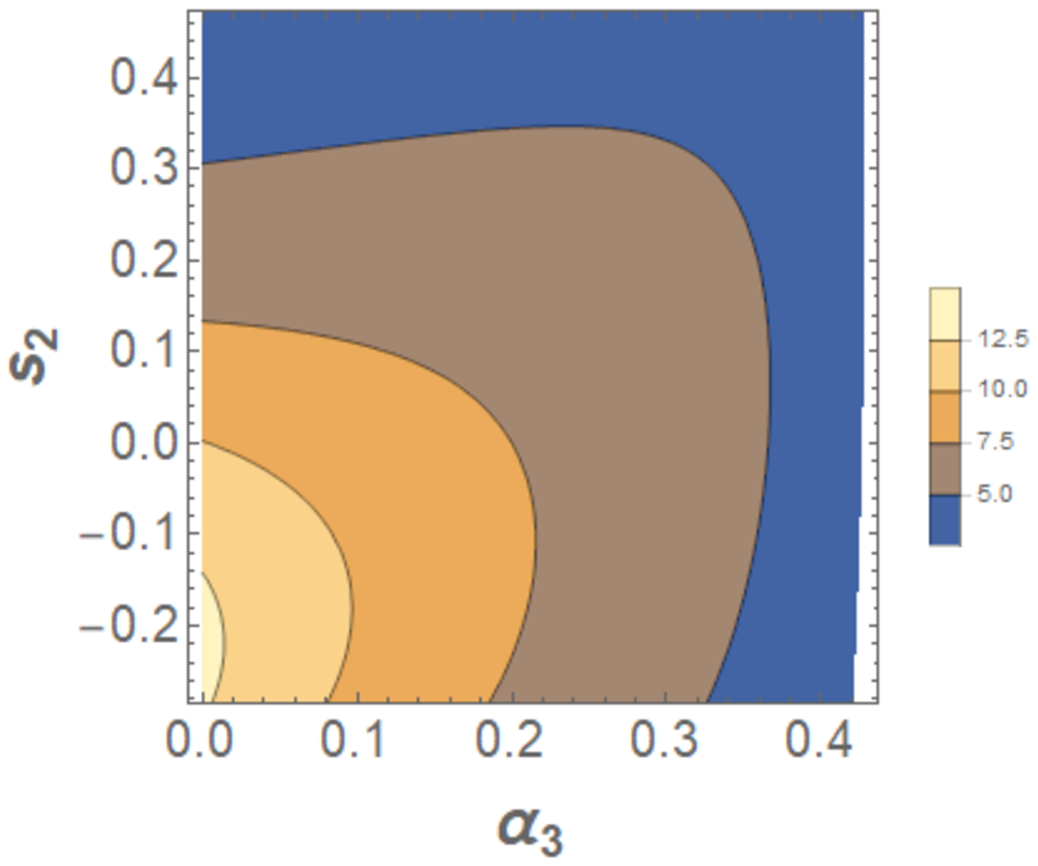}
	}
	\subfigure[{}]{
		\includegraphics[width=0.31\textwidth]{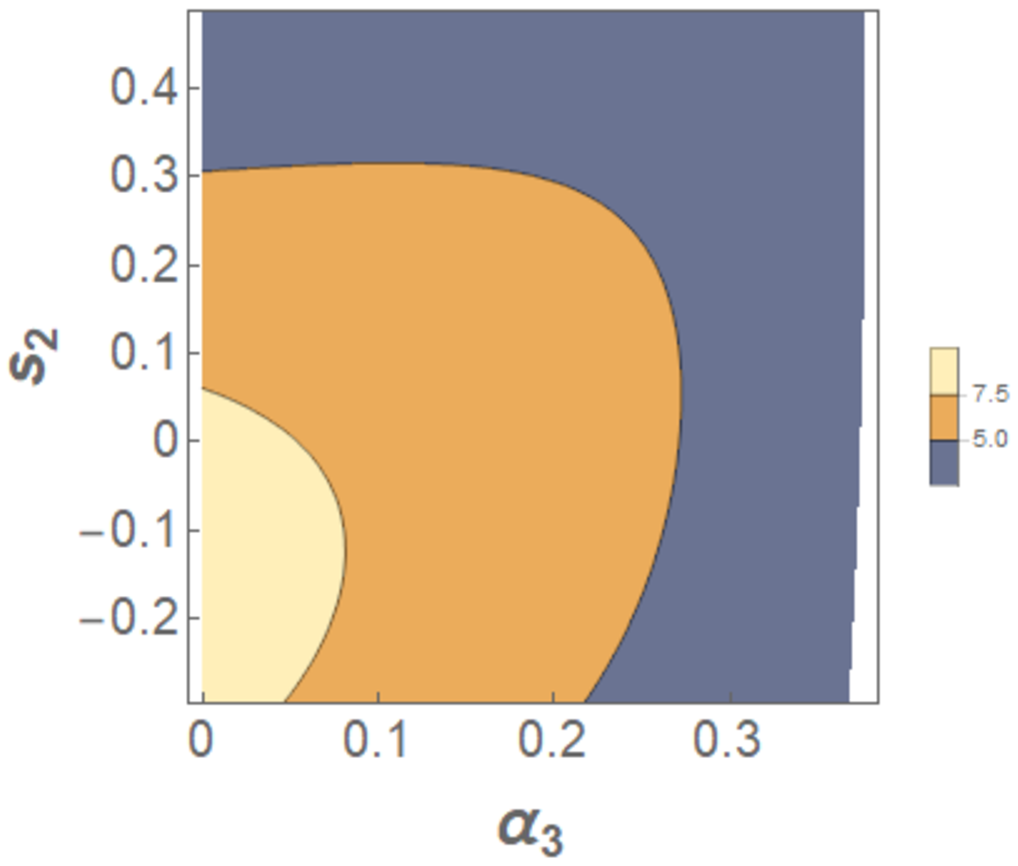}
	}
	\caption{The contour maps of $E_3$ in terms of $\alpha_3$ and $s_2$ with $s_1=0$. (a)($b=0$), (b)($b=0.1$) and (c)($b=0.2$) show that $\alpha_3 \rightarrow 0$ give larger value of $E_3$ when $s_2$ is small enough.}
	\label{MMMfig:contourmapsofE3} 
\end{figure}

From the Eq. \eqref{MMMformulaE3}, we find that the expression of $E_3$ decoupled with the parameters $\xi$ and $\beta_3$. So we analyze the maximal value of $E_3$  with the remaining parameters for different values of $b$. Note that
Fig. \ref{fig:conditionofthespin} shows that the spin magnitude $s_1$ close to zero for larger value of  $E_3$. So we first assume $s_1=0$ in order to find the relation of $E_3$ and $\alpha_3$. The contour maps of $E_3$ in terms of $\alpha_3$ and $s_2$  showed in Fig. \ref{MMMfig:contourmapsofE3}.
From the Fig. \ref{MMMfig:contourmapsofE3}, we know  that the largest efficiency can found with $\alpha_3 \rightarrow 0^+$. Therefore, we set $\alpha_3=0^+$ to calculate the corresponding maximal efficiency.
\begin{figure}[!htb]
	\centering
	\subfigure[{}]{
		\includegraphics[width=0.31\textwidth]{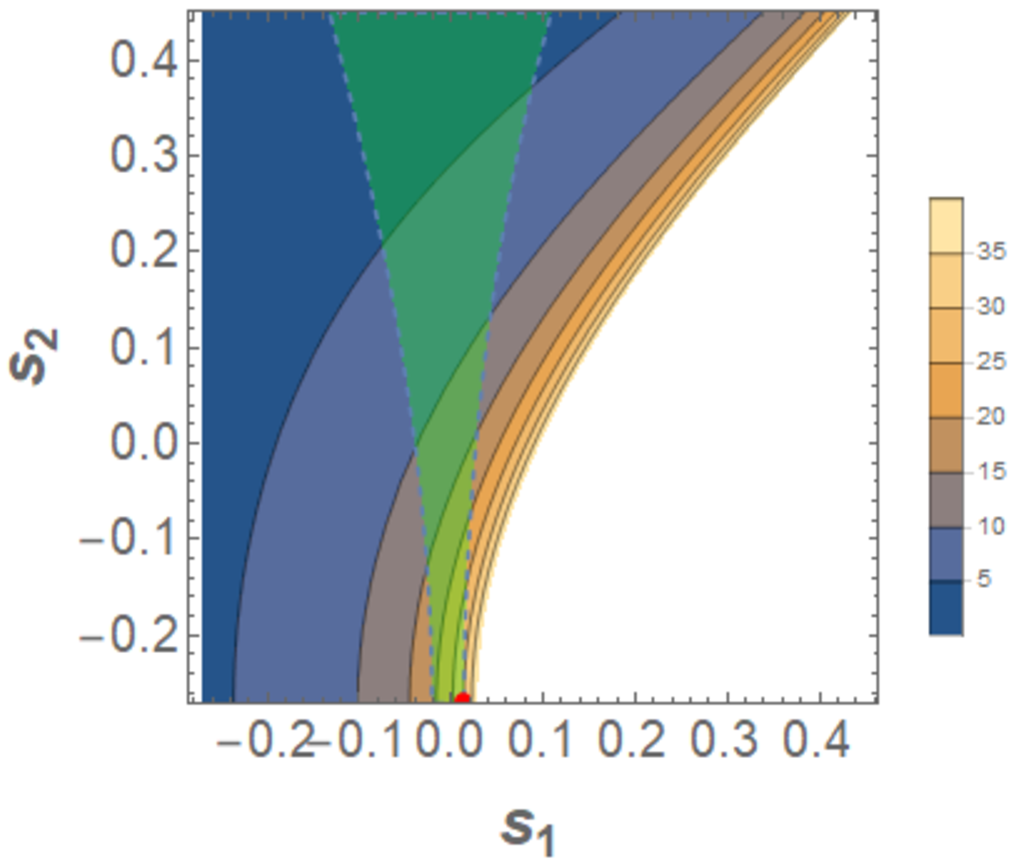}
	}
	\subfigure[{}]{
		\includegraphics[width=0.31\textwidth]{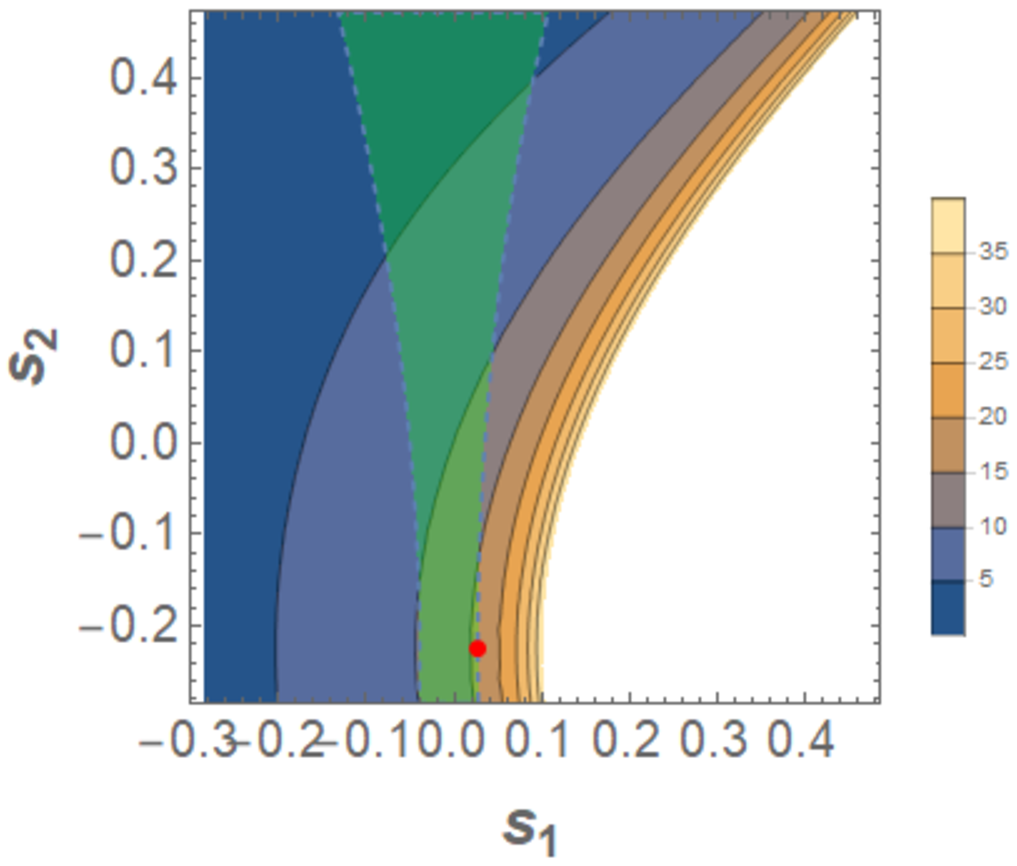}
	}
	\subfigure[{}]{
		\includegraphics[width=0.31\textwidth]{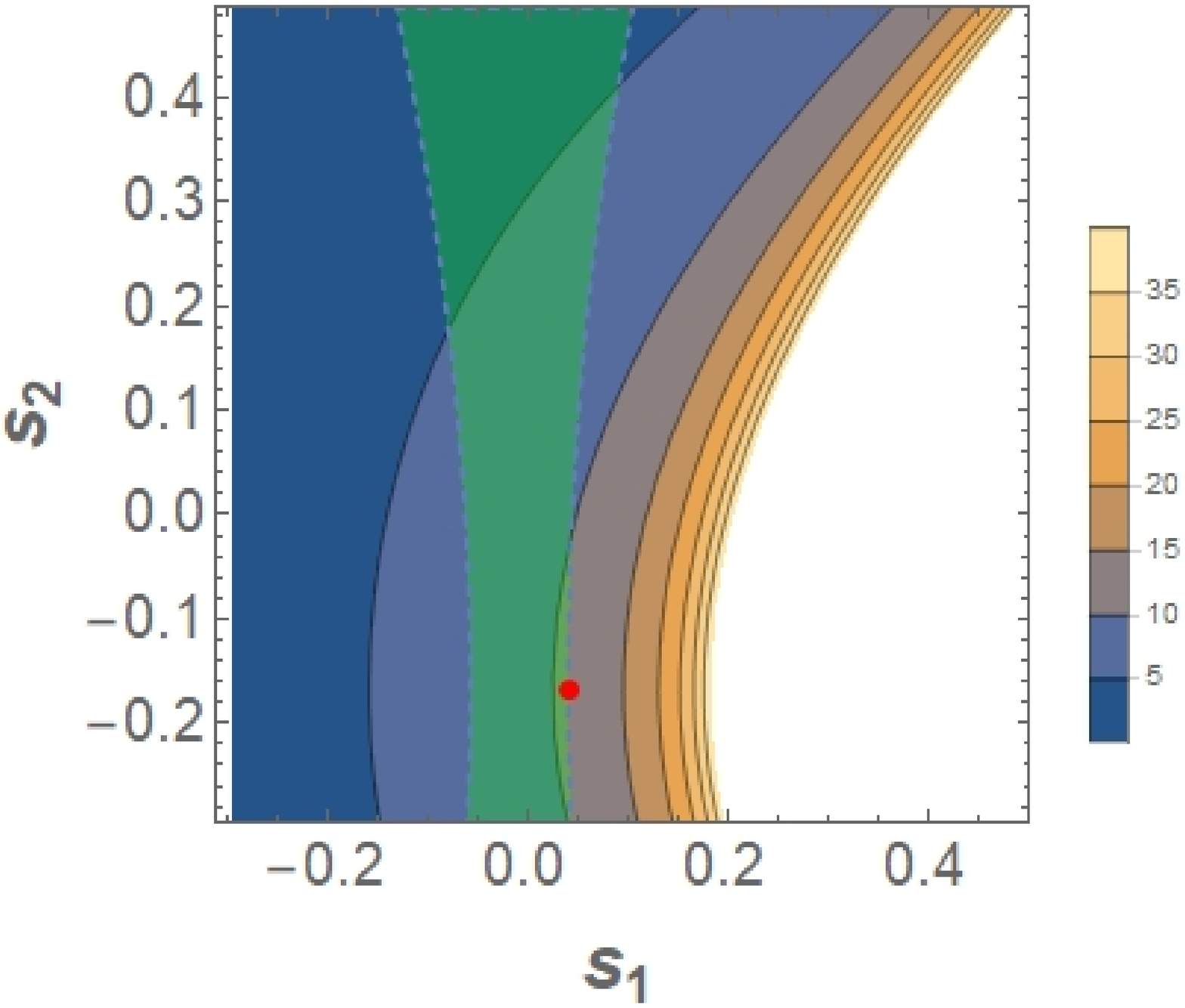}
	}
	\caption{The contour map of $E_3$ in terms of $s_1$ and $s_2$. The time-like condition for the particle 3 orbit is satisfied in the green shaded region. The maximum value of $E_3$ is obtained at the red point.	(a)when $b=0$, $\eta_{max}=E_{3max}/2\approx 15.01$ at $(s_1=0.01378, s_2=-0.2679)$; (b) when $b=0.1$, $\eta_{max}=E_{3max}/2\approx 7.964$ at $(s_1=0.02694, s_2=-0.2253)$; (c)when $b=0.2$, $\eta_{max}=E_{3max}/2\approx 5.378$  at $(s_1=0.04076, s_2=-0.1680)$}
	\label{fig:E3ofMMM} 
\end{figure}

In Fig. \ref{fig:E3ofMMM}, the contour map of $E_3$ in terms of $s_1$ and $s_2$ is showed. The maximal value of $E_3$ is labeled with the red point.

\begin{figure}[!htb]
	\centering
	\subfigure[{}]{
		\label{fig:E2ofMMM}
		\includegraphics[width=0.4\textwidth]{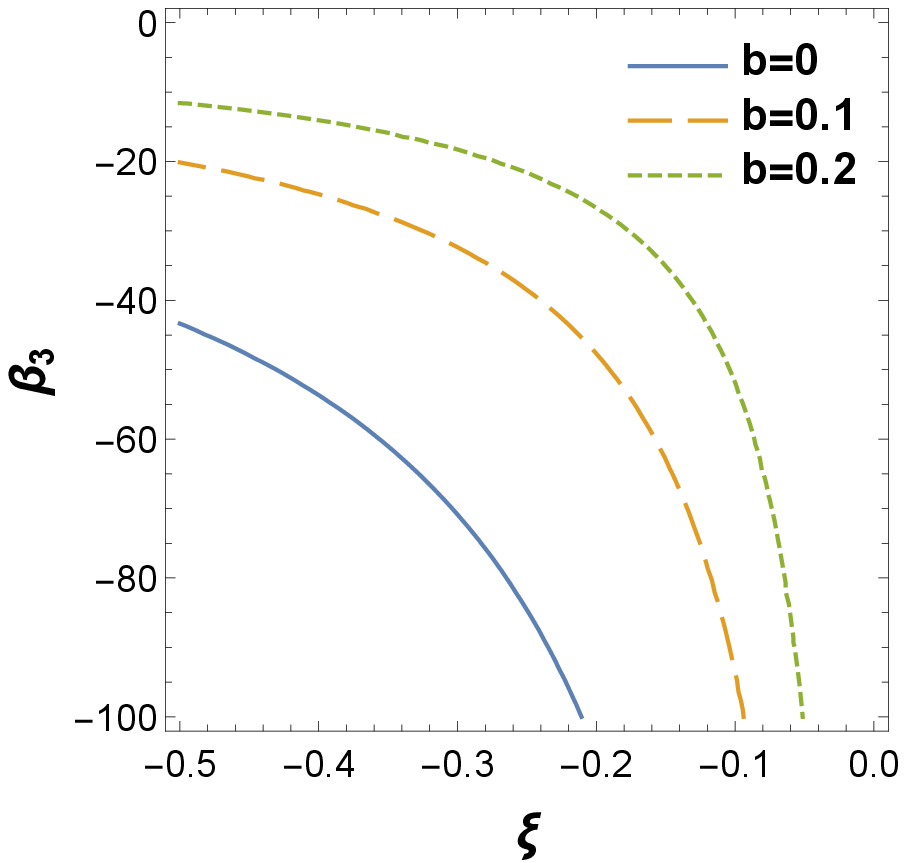}
	}
	\subfigure[{}]{
		\label{fig:bandxiofMMM}
		\includegraphics[width=0.4\textwidth]{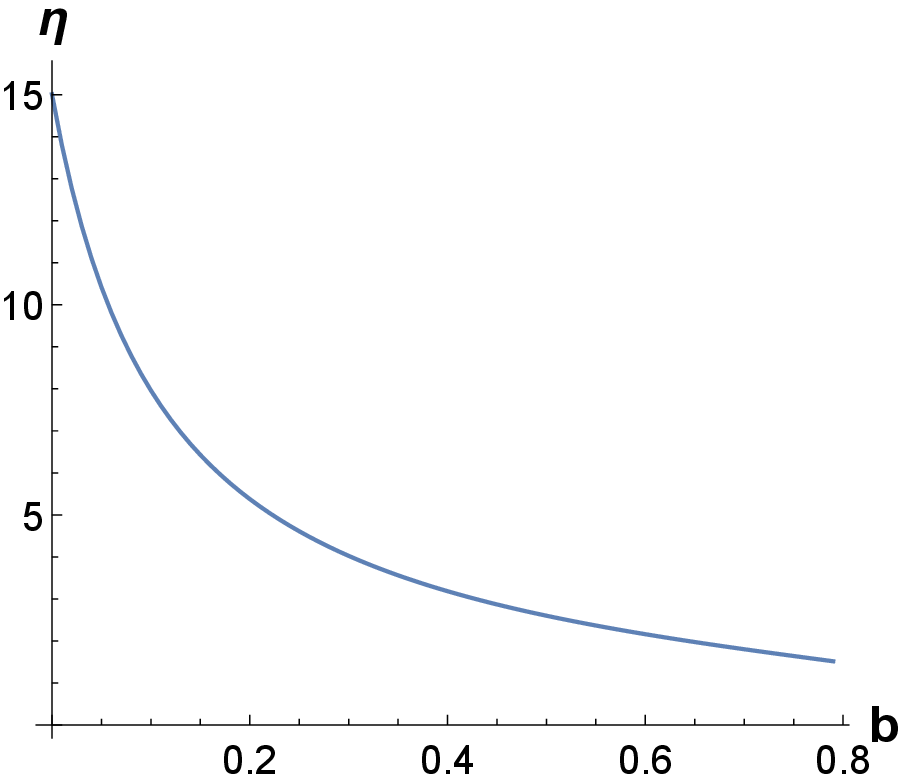}
	}
	\caption{(a) The relation between $\xi$ and $\beta_3$ for different value of $b$ when $E_2=1$. The other parameters are chosen for giving the maximal value of $E_3$. When $b=0$, $s_1 = 0.01378$ and $s_2 = -0.2679$;  When $b=0.1$, $s_1 = 0.0269$ and $s_2 = -0.225$; When $b=0.2$, $s_1 = 0.0408$ and $s_2 =-0.168$; (b)The relation between maximum efficiency $\eta_{max}=E_{3max}/2$ and $b$.}
\end{figure}

Note that $E_2 \geq 1$ if the particle 2 falling from infinity, if $E_2=1$ is possible, we find that the maximal
value of $E_3$ gives the maximal efficiency. Note that $E_3$ is decoupled with parameters $\beta_3$ and $\xi$. So our target is equivalent to find $E_2=1$ with some admissible values of $\beta_3$ and $\xi$. In Fig. \ref{fig:conditionofxi}, we already have the constraint on $\xi$, that is, $0 > \xi \ge -0.5 > \xi_{min}$ for different values of $s$ and $b$. For such constrained $\xi$, the  relation between $\beta_3$ and $\xi$ which gives $E_2=1$ can be found in Fig. \ref{fig:E2ofMMM}.

Hence the maximum efficiency is given by $\eta_{max}=E_{3max}/2$. The Fig. \ref{fig:bandxiofMMM} shows the maximum efficiency $\eta_{max}$ with different $b$. We found that
the efficiency $\eta_{max}$  decreases with the increase of $b$. While when $b=0$ which corresponds the Kerr case, our results is the same as the previous results\cite{PhysRevD.98.064027}.

\subsection{Maximal Efficiency in Case [B] MPM}
For the case[B], as the same with subsection, we assume that the mass of massive particles are all equal to $m$ and the massless particles are nonspinning.
The equations of conservation law \eqref{conservations} and \eqref{conservationp} reduce to
\begin{eqnarray}
s_1 &=& s_3\\
p_1^{(1)}+p_2^{(1)} &=& p_3^{(1)}+p_4^{(1)}
\end{eqnarray}
 The radial component of the 4-momentum of massless particle can be calculated from the Eqs. \eqref{MMMformulamotionequationp0} - \eqref{MMMformulamotionequationpD}, and \eqref{normalizationcondition} as follows
\begin{eqnarray}
\label{masslessparticlesp}
p^{(1)}=\sigma  \sqrt{\frac{(E (b+r+1)-J) \left(E \left((3 b-1) r+2 (b-1)^2+r^2\right)+J (2 b+r-2)\right)}{(b+r-1)^2 (2 b+r)}}
\end{eqnarray}
So the expression of $f_{22}$, $f_{23}$, $f_{42}$, and $f_{43}$ can write in an explicit way:
\begin{small}
\begin{eqnarray}
f_{22}&=&-\frac{2 E_2 (b \xi +b+2 \xi +1)}{(b+1) \sqrt{1-b^2}}\\
f_{23}&=&-\frac{E_2 \left(4 (2 b-1) \xi ^2+8 b (b+1) \xi +(b+1)^4\right)}{4 (b+1)^2 \sqrt{1-b^2} \xi }\\
f_{42}&=&-\frac{2 (b+1) (E_1+E_2+(\alpha_3-1) E_3)+2 (b+2) E_2 \xi }{(b+1) \sqrt{1-b^2}}\\
f_{43}&=&-\frac{1}{4 (b+1)^2 \sqrt{1-b^2} E_2 \xi } \bigg( 8 (b+1) E_2 \xi  \left(b \left(E_1+E_2+E_3 (-\alpha _3+\beta _3-1)\right)+E_3 \left(\beta_3-2 \alpha_3 \right) \right)
\notag \\
&&+(b+1)^4 (E_1 + E_2 -E_3)^2+4 (2 b-1) E_2^2 \xi ^2 \bigg)
\end{eqnarray}
\end{small}
Note that the radial component of the 4-momentum of massive particle do not change through the collision process.
As in case [A], we finally get the detail expression for $E_3$ and $E_2$ respectively.
\begin{eqnarray}
E_3&=&\frac{\sqrt{\mathcal{B}_1^2-\mathcal{A}_1 \mathcal{C}_1}+\mathcal{B}_1}{\mathcal{A}_1}\bigg|_{s_2=0}\\
E_2&=&\frac{(b-1) (E_1-E_3)^2}{\mathcal{P}_1} \bigg|_{s_2=0}
\end{eqnarray}
where $\mathcal{A}_1$, $\mathcal{B}_1$, $\mathcal{C}_1$ and $\mathcal{P}_1$ are given by Eqs. \eqref{MMMformulaABC}  and \eqref{MMMformulaP} with $s_2=0$.

\subsubsection{Efficiency}
On the one hand, when the value of $E_1$ is given, the maximal efficiency $\eta_{max}$ would be reached  with minimum value of $E_2$ and maximal value of $E_3$. On the other hand, we consider particle 1 and particle 2 falling from infinity, we obtain the constrains of $E_1 \ge 1$ and $E_2 \ge 0$. Without loss generality, we again normalize $E_1$ to unity ($E_1=1$) as last subsection£¬ and then analyze $E_3$ and $E_2$ which are directly associated to the maximal efficiency.
\begin{figure}[!htb]
	\centering
	\subfigure[{}]{
		
		\includegraphics[width=0.31\textwidth]{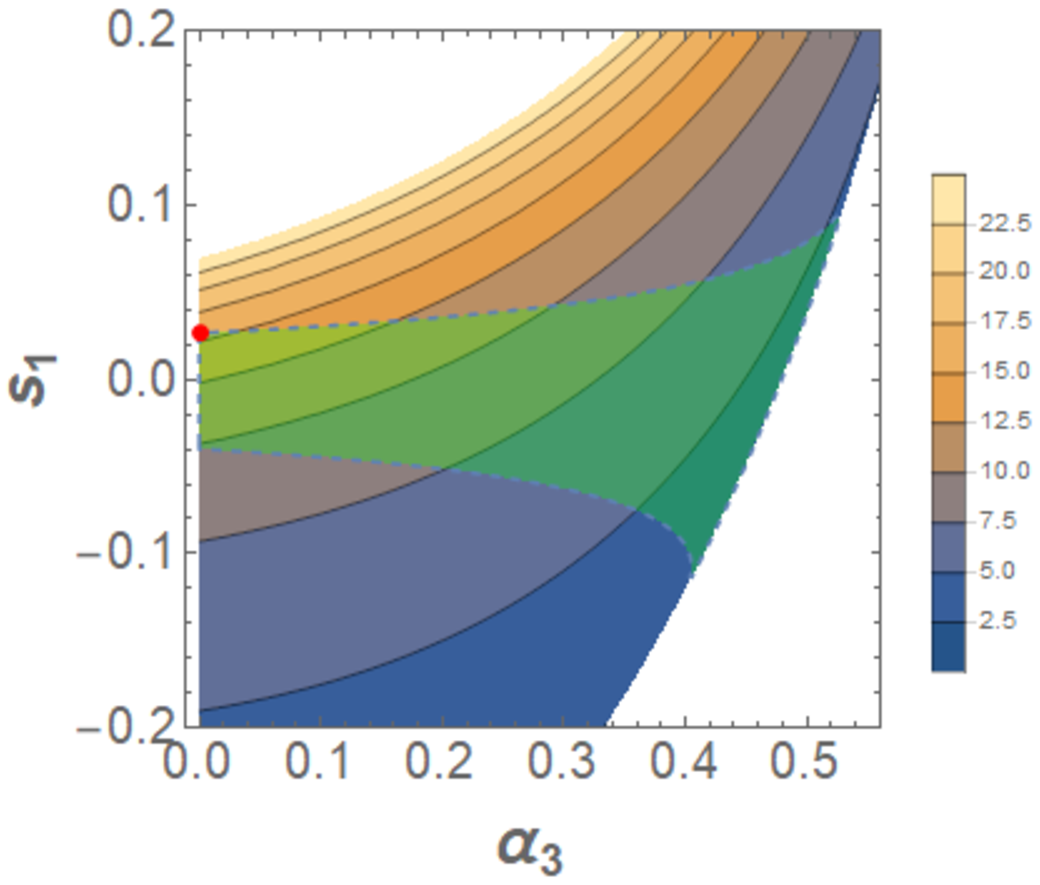}
	}
	\subfigure[{}]{
		\label{fig:E3ofMPMb}
		\includegraphics[width=0.31\textwidth]{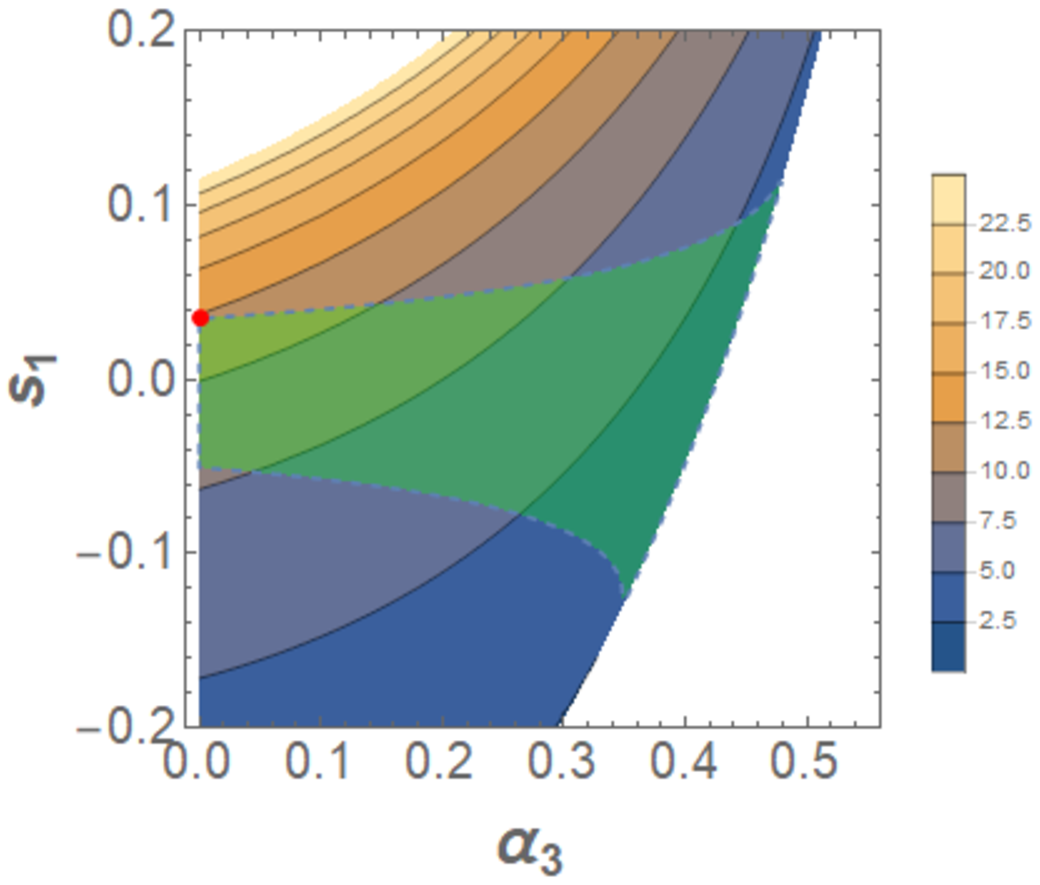}
	}
	\subfigure[{}]{
		\includegraphics[width=0.31\textwidth]{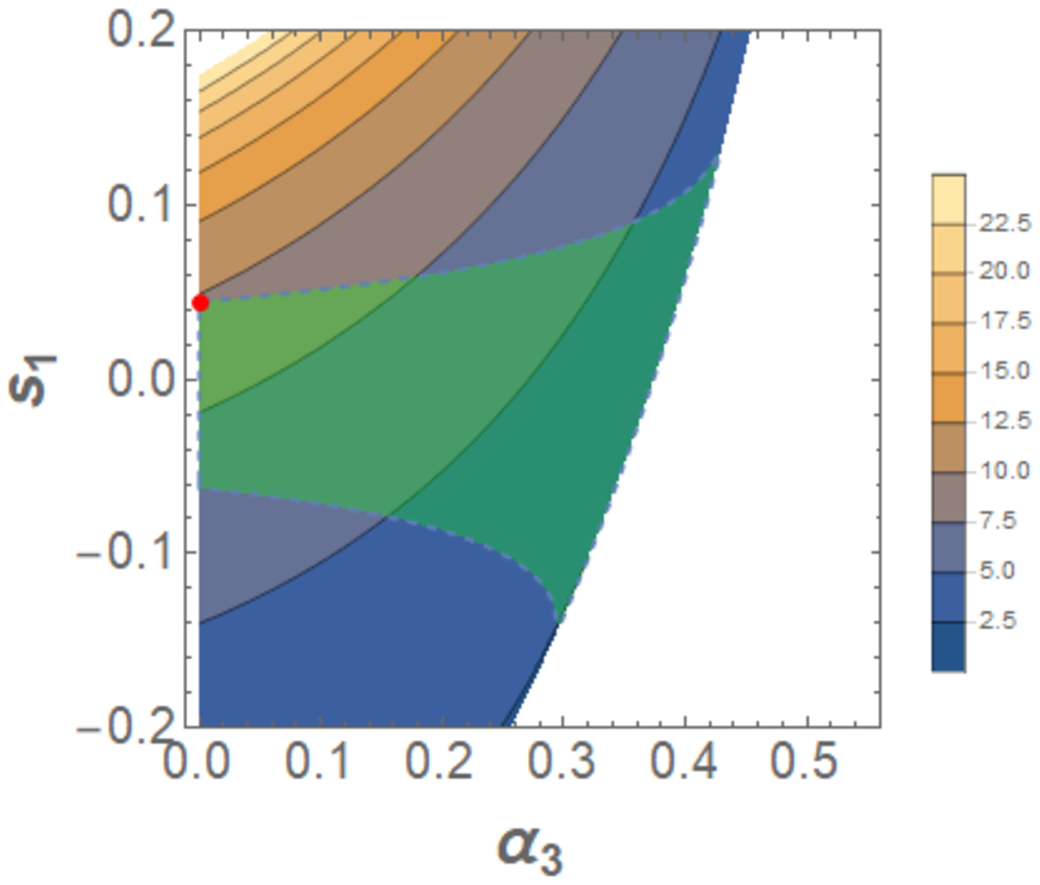}
	}
	\caption{The contour map of $E_3$ in terms of $\alpha_3$ and $s_1$. The time-like condition for the particle 3 is satisfied in the light-green shaded region with different $b$. The maximum value of $E_3$ is obtained at the red point. (a)when $b=0$, $E_{3max}\approx 15.6350$; (b) when $b=0.1$, $E_{3max}\approx 12.2977$; (c)when $b=0.2$, $E_{3max}\approx 9.7977$}
	\label{fig:E3ofMPM} 
\end{figure}

Fig. \ref{fig:E3ofMPM} shows the contour map of $E_3$ in terms of $\alpha_3$ and $s_1$.  The maximum value of $E_3$ is given at the red point.

If $E_2 \rightarrow 0$ can be achieved, it certainly gives the minimal value of $E_2$ and thus the maximal efficiency can be simply given by $\eta_{max} = E_{3max}$.
Hence it is important to analyze
whether $E_2 \rightarrow 0$ is possible or not. From Eq. \eqref{MMMformulaP}, we obtain the asymptotic expression of $\mathcal{P}$ as
\begin{eqnarray}
\mathcal{P} \approx 4 (b+1) E_3 \beta_3 \xi  \left(\frac{\sqrt{1-b^2} E_3 h_{61}(s_1,b,0)}{f_1(s_1,b)^2 k_1(E_3,s_1,b,0)}-2\right)\label{asymptoticP}
\end{eqnarray}
Eq. \eqref{asymptoticP} tells us that $E_2 \rightarrow 0^+$, if $\beta_3 \xi \rightarrow +\infty$.  For example,  the value of parameters at red point in Fig. \ref{fig:E3ofMPMb} are $b=0.1$, $\sigma_1=1$,  $\sigma_3=-1$,  $\alpha_3=0$,  $s_1=0.03513$,  $s_2=0$,  $E_1=1$,  $E_3=12.2977$. So the detail expression of $E_2$ can be rewritten as:
\begin{eqnarray}
E_2 =\frac{7.27275}{ \beta_3  \xi +1.74978 \xi +1.28748}.
\end{eqnarray}
which means $E_2 \rightarrow 0^+$ can be realized $\beta_3 \xi \rightarrow \infty$ for the case of $b=0.1$.

Hence by employing formula $\eta_{max} = E_{3max}$, we found that the efficiency $\eta_{max}$  decreases with the increase of $b$. While $b=0$ which corresponds the Kerr case, our results is again the same as the previous results\cite{PhysRevD.98.064027}.

\subsection{Maximal Efficiency in Case [C] PMP}
 Now we come to the last case, which is the Compton scattering. The radial components of 4-momenta of  massless particles  have already been given in Eq. \eqref{masslessparticlesp}.
So we can write the coefficients $f_{12}$,  $f_{13}$,  $f_{32}$ and $f_{33}$ in terms of energy as:
\begin{small}
\begin{eqnarray}
f_{12}&=&\sqrt{\frac{b+3}{b+1}} E_1 \sigma_1\\
\label{PMPformulaf13}
f_{13}&=&\frac{(b-1) E_1 \sigma_1}{(b+1)^{3/2} \sqrt{b+3}}\\
f_{32}&=& E_3 \sigma_3 \sqrt{\frac{-4 \alpha_3^2+8 \alpha_3+b^2+2 b-3}{b^2-1}}\\
\label{PMPformulaf33}
f_{33}&=&\frac{E_3 \sigma_3 \left(8 \alpha_3^2-8 \alpha_3+b^2+4 \alpha_3^2 b-4 (\alpha_3-1) (b+1) \beta_3-2 b+1\right)}{(b+1) \sqrt{\left(1-b^2\right) (3-2 \alpha_3+b) (1-2 \alpha_3-b)}}
\end{eqnarray}
\end{small}
From the conservation of the radial components of the 4-momenta,  we find
\begin{eqnarray}
E_3 =\mathcal{S} E_1
\end{eqnarray}
where the amplification factor $\mathcal{S}$ is given by
\begin{eqnarray}
\mathcal{S}=\frac{\sqrt{1-b} \left(\sqrt{b+3} \sigma_1 f_1(s_2, b)+\sqrt{b+1} h_{71}(s_2, b)\right)}{\sigma_3 \sqrt{(3-2 \alpha_3+b) (1-2 \alpha_3-b)} f_1(s_2, b)-\sqrt{1-b^2} h_{72}(s_2, b, \alpha_3)}
\end{eqnarray}
and
\begin{eqnarray}
\label{PMPformulaE2}
E_2=-\frac{(b-1) (E_1-E_3)^2 f_1(s_2, b) h_{81}(s_2, b)}{\mathcal{P}_3};
\end{eqnarray}
where $\mathcal{P}_3$ keeps the same form of $\mathcal{P}_1$ given by Eq. \eqref{MMMformulaP} by replaceing $f_{13}$ and $f_{33}$ with Eqs. \eqref{PMPformulaf13} and \eqref{PMPformulaf33}

\subsubsection{Efficiency}

It is easy to see in Compton scattering, the efficiency $\eta$ is defined as:
\begin{eqnarray}
\eta=-\frac{\mathcal{S}}{1+E_2/E_1};
\end{eqnarray}
again, we consider massless particle 1 and massive particle 2 falling from infinity,  we assume the constrains of $E_1 \ge 0$ and $E_2 \ge 1$ and obtain that the maximal value of $\mathcal{S}$ and the minimal value of $E_2/E_1$ gives the maximal efficiency. First,  we can easily find that the ratio $E_2/E_1$ doesn't depends on the $E_1$ and $E_2$,  but rather depends on the parameters $\alpha_3$,  $\beta_3$,  $\xi$,  $s_2$ and $b$. From Eq. \eqref{PMPformulaE2}, the asymptotic expression of $E_2/E_1$  behaves
\begin{small}
\begin{eqnarray}
\label{MMMformulaE2/E1}
\frac{E_2}{E_1} \approx \frac{(b-1) (\mathcal{S}-1)^2 h_{81}(s_2, b)\sqrt{(-2 \alpha_3+b+3) (1-b-2 \alpha_3)}}{8 \beta_3 \xi  \mathcal{S} f_2(s_2, b) \left(2 (\alpha_3-1) f_1(s_2, b)-(b+1) f_2(s_2, b) \sqrt{(-2 \alpha_3+b+3) (1-b-2 \alpha_3)}\right)}
\end{eqnarray}
\end{small}

Note that the particle 2 is massive and can reach the horizon, therefore the constraint on $\xi$ keeps the same form as in previous section, namely, $\xi_{min}<\xi<0$. With this parameter space, a direct calculation shows that $\mathcal{S} \neq 0$. From Eq. \eqref{MMMformulaE2/E1}, we can see that if denominator of  the equation is not equal to zero, the condition $E_2/E_1 \rightarrow 0$ can be archived when $\beta_3 \xi \rightarrow \infty$.
Thus the maximal energy contraction efficiency is $\eta_{max}=\mathcal{S}_{max}$.
\begin{figure}[!htb]
	\centering
	\subfigure[{}]{
		
		\includegraphics[width=0.31\textwidth]{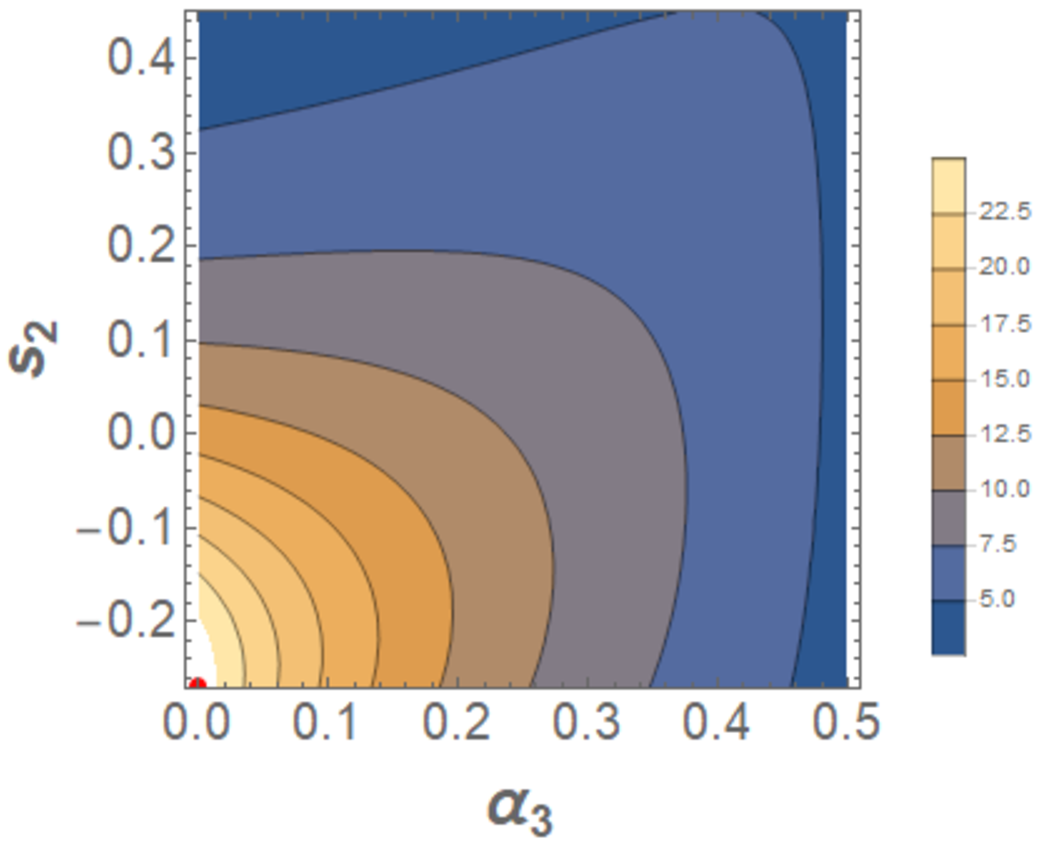}
	}
	\subfigure[{}]{
		\includegraphics[width=0.31\textwidth]{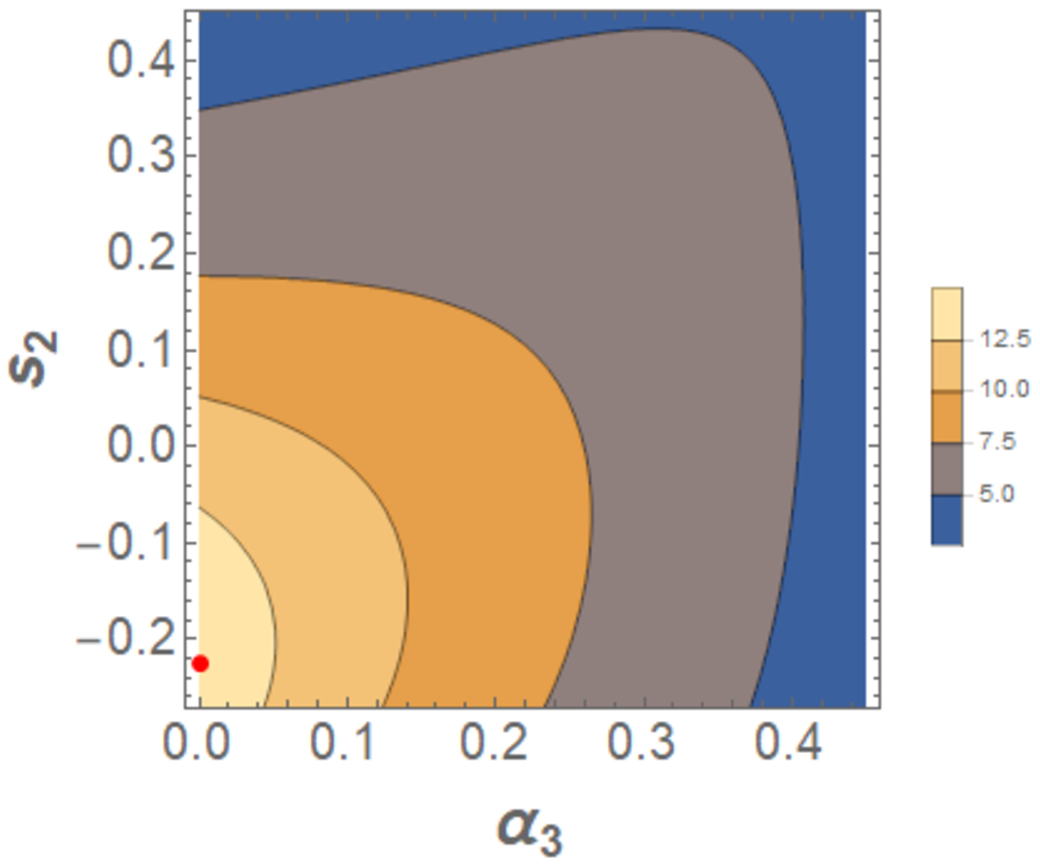}
	}
	\subfigure[{}]{
		\includegraphics[width=0.31\textwidth]{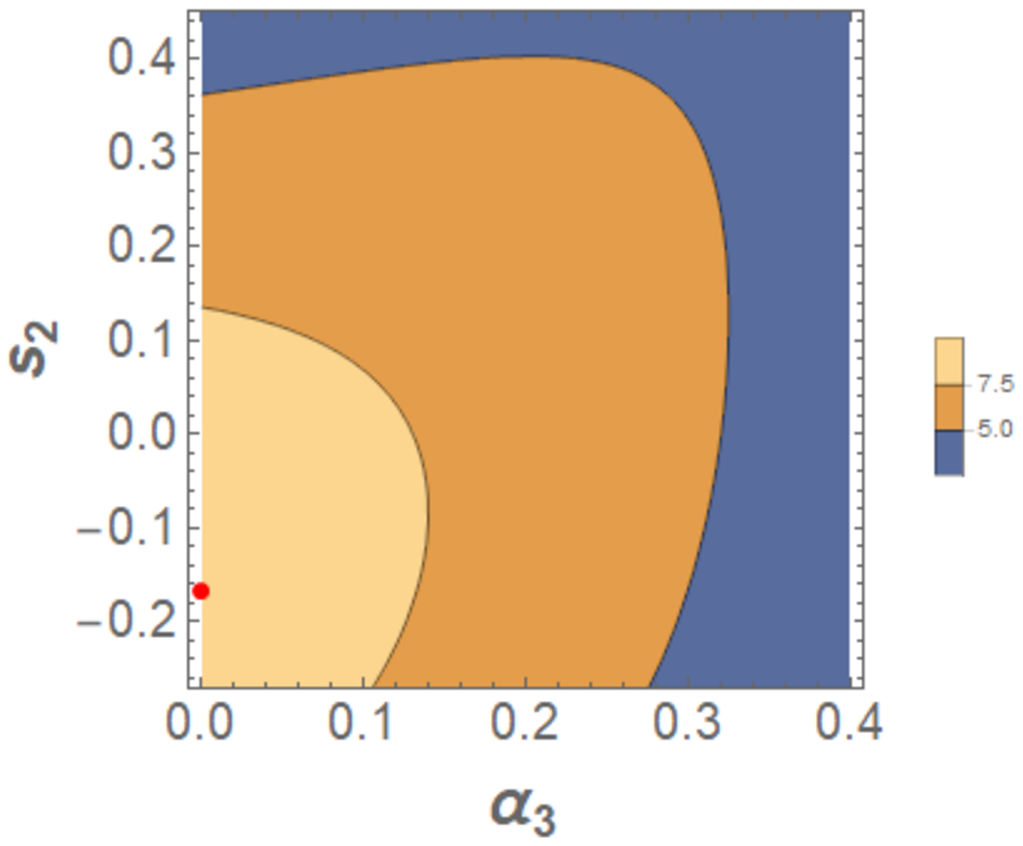}
	}
	\caption{The contour map of $\mathcal{S}$ in terms of $\alpha_3$ and $s_2$ for different values of $b$. The maximum value of $E_3$ is labeled by the red point. (a)when $b=0$,  $\eta_{max}=\mathcal{S}_{max}\approx 26.8564$ with $s_2=-0.2679$; (b) when $b=0.1$,  $\eta_{max}=\mathcal{S}_{max}\approx 14.4513$ with $s_2=-0.2253$; (c)when $b=0.2$,  $\eta_{max}=\mathcal{S}_{max}\approx 9.7977$ with $s_2=-0.1680$}
	\label{fig:PMPE3} 
\end{figure}

The Fig. \ref{fig:PMPE3} shows the maximum value of $E_3$  with the red point in the contour map of $\mathcal{S}$ in terms of $\alpha_3$ and $s_2$ for different values of $b$. The figure shows the maximum efficiency $\eta_{max} = \mathcal{S}_{max}$ decreases when $b$ increases.

\section{CONCLUSIONS}
In this paper,  we study the collision of two uncharged spinning particles around an extreme Kerr-Sen black hole and calculate the maximal efficiency of the energy extraction from the black hole. We consider the  particles freely falling from infinity to the Kerr-Sen black hole. The Kerr-Sen spacetime is determined by three parameters, which are mass $M$,  angular momentum $a$,  and charge $Q$($b=Q^2/2M$). It reduces to a Kerr black hole when the parameter $b=0$ and all our results coming back to the Kerr case\cite{PhysRevD.98.064027} when $b=0$. We viewed this as
a consistent check.

In this paper, we consider three types of collision, the first one is the MMM case[A],  we obtain that the maximum efficiency is given by $\eta_{max}=E_{3max}/2$ and  decreases monotonously  with the increase of $b$.
Then,  in the MPM case[B],  we obtain the maximum efficiency $\eta_{max} = E_{3max}$ and decreases monotonously  with the increase of $b$.
Finally,  in the PMP case[C],  we get the maximum efficiency $\eta_{max} = \mathcal{S}_{max}$  which decreases when the $b$ increases. All our results can reduce to the Kerr situation \cite{PhysRevD.98.064027} when $b=0$. The Compton scatting and inverse Compton scatting of spinless particle in Kerr background is discussed in \cite{PhysRevD.86.024027}, and our results
shows when the spin take into account, the maximum efficiency can be greatly improved.

In summarize, for extreme Kerr-Sen black hole, decrease the charge parameter $b=Q^2/2M$ always increase the maximum efficiency of energy extraction.

\begin{acknowledgements}
This work is supported by NSFC with No.11775082. The authors could like to thank prof. Kazumasa Okabayashi for helpful discussion.

\end{acknowledgements}

\section{APPENDIX}
\begin{eqnarray}
\mathcal{U}&=&-a^2 b^4 \Delta  r^2 s^4 (C_{t3}-Br C_{t4})^2+4 a b^2 C_{ta2} r^3 s^2 (Br C_{t2}+C_{t1}) (C_{t3}- Br C_{t4} ) \notag\\
&&+2 f_{v2} r (C_{t3}-Br C_{t4}) \Big(a b^2 s^2 (Br C_{t2}+C_{t1}) +C_{ta2} r^3 (C_{t3}-Br C_{t4})\Big) \notag\\
&&+ 2 b^2 r s^2 (Br C_{t2}+C_{t1} ) \left(a b^2 \Delta  s^2 (C_{t3}-Br C_{t4})-C_{ta2} r (Br C_{t2}+ C_{t1})\right) \notag\\
&&-b^4 \Delta  s^4 (Br C_{t2} + C_{t1})^2+ f_{v2} ^2 r^2 (C_{t3}-Br C_{t4})^2+a^2 b^4 s^4 (Br C_{t2}+C_{t1})^2\\
C_{ta2}&=&\mathcal{F}/r^2;\\
C_{t1}&=&\sqrt{\Sigma} \left(a^2 r (2 b+r)^2+a s ((b+r) (2 b+r)+r) \sqrt{\Sigma}+ (2 b+r) \Sigma^2 \right);\\
C_{t2}&=&\sqrt{\Sigma} \left(-a r (2 b+r)^2-r s \sqrt{\Sigma }\right);\\
C_{t3}&=&(2 b+r)^2 \left(a \sqrt{\Sigma }+s (b+r)\right);\\
C_{t4}&=&\sqrt{\Sigma } (2 b+r)^2;
\end{eqnarray}
\begin{eqnarray}
f_1(s, b)&=&-b^4-2 b^3+2 b-s^2+1;\\
f_2(s, b)&=&b^3-\sqrt{1-b^2} s+b^2-b-1;\\
k_1(E, s, b, \alpha)&=&\sqrt{E^2 k_{12}(s, b, \alpha)-f_1(s, b)^2};\\
k_2(s, b, \xi ) &=& -\xi  \Big(2 \left(b^2-3 b+1\right) \sqrt{1-b^2} (b+1) s^2 -2 (b-1) (b+2) \sqrt{1-b^2} (b+1)^4 \notag \\
&&+2 (b-1) b s^3+2 (b-3) (b-1) (b+1)^3 s\Big)-(b+1) \Big(\left(b^2+2 b-1\right) s^3\notag \\
&&-2 \sqrt{1-b^2} (b+1) s^2+(b-1) \left(b^2+2 b-1\right) (b+1)^3 s \notag \\
&&-2 (b-1) \sqrt{1-b^2} (b+1)^4\Big);
\end{eqnarray}
\begin{eqnarray}
k_3(s, b)&=&(b-1) (b+1) \Big(-(b-1) (3 b-7) \sqrt{1-b^2} (b+1)^4 s+\left(b^5+3 b^4+7 b+1\right) s^4\notag \\
&&+(b-1)^3 (b+1)^6 -\left(2 b^3+2 b^2-3 b+11\right) \sqrt{1-b^2} (b+1) s^3 \notag\\
&&+(b-1) \left(2 b^3-2 b^2-13 b+9\right) (b+1)^3 s^2\Big);\\
k_{12}(s,  b,  \alpha) &=& (b+1)^2 \bigg(\alpha ^2 \left(8 \sqrt{1-b^2} (b+1) s-4 (b-1) (b+1)^3+4 s^2\right) +4 \alpha  \Big(\big(b^2+2 b-1\big)\notag \\
&& s^2+(b-1) (b+3) \sqrt{1-b^2} (b+1) s+2 (b-1) (b+1)^3\Big)  +(b-1) (b+3) \notag\\
&& \Big( b (b+2) s^2 +(b-1) (b+1)^3 -2 \big((b+1) \sqrt{1-b^2}\big) s \Big)\bigg)\\
h_{41}(s, b, \xi ) &=& (b+1) \bigg(4 (b-1) \xi  \Big(2 (b-1) \left(b^2-2 b+2\right) \sqrt{1-b^2} (b+1)^4 s\notag \\
&&+\big(b^4-2 b^2+2 b-1\big) s^4 +\left(b^4-2 b^2+8 b-5\right) \sqrt{1-b^2} (b+1) s^3\notag \\
&&+2 (b-2)^2 (b-1) (b+1)^3 s^2 +2 (b-1)^2 b (b+1)^6 \Big) f_1(s, b)-(b+1)^2 \left(b^2-1\right)  \notag \\
&&\big(b^4+2 b^3 +2 b \big(\sqrt{1-b^2} s-1\big)+2 \sqrt{1-b^2} s-s^2-1\big) f_1(s,b)^2-4 (b-1) \xi ^2 \notag \\
&&\Big((b-1)^3 (2 b-1) (b+1)^8 +(b-1)^2 \left(7 b^2-12 b+8\right) \sqrt{1-b^2} (b+1)^6 s\notag \\
&&+2 (b-1) b^2 s^6 +2 (b-1) \big(6 b^3-16 b^2+16 b-13\big) \sqrt{1-b^2} (b+1)^3 s^3 \notag \\
&&-2 (b-1)^2\left(2 b^3-12 b^2+14 b-11\right) (b+1)^5 s^2 -(b-1) \big(2 b^4-12 b^3\notag \\
&&+18 b^2-18 b+13\big) (b+1)^2 s^4 +\big(4 b^4-4 b^3+b^2-4 b+2\big) \sqrt{1-b^2} s^5 \Big)\bigg)\\
h_{61}(s,  b,  \alpha ) &=& (b+1) \Big(2 (b-1) (b+3) \sqrt{1-b^2} (b+1)^2 s^3+2 (b-1)^2 (b+3) \sqrt{1-b^2} (b+1)^5 s \notag \\
&&+4 \alpha  (b+1) \Big(2 \sqrt{1-b^2} (b+1) s^3+2 (b-1) \sqrt{1-b^2} (b+1)^4 s-(b-1)^2 (b+1)^6 \notag\\
&&+s^4\Big)+2 \left(b^3+3 b^2+b-1\right) s^4+2 (b-1) (b+1)^6 s^2+4 (b-1)^2 (b+1)^7 \Big)\\
h_{62}(s,  b,  \alpha)&=&\alpha h_{621}(s, b)+\alpha ^2 h_{622}(s, b)\\
h_{621}(s,  b) &=& -2 (b+1) (b-1) \Big(  -(b-1) (b (2 b+3)-6) \sqrt{1-b^2} (b+1) s^3 \notag \\
&&-(b+1)^3 (b (b (b+3)-21)+13) s^2+(b-1) (b (b+3)+12) \sqrt{1-b^2} (b+1)^4 s  \notag \\
&&+4 (b-1) (b+1)^6+(-b (b+2) (2 b-1)-1) s^4 \Big)\\
h_{622}(s,  b) &=& 4 (b+1) \Big( -5 \left((b-1) (b+1)^4 \sqrt{1-b^2}\right) s+\left(2 b^3-2 b^2-3 b+1\right) \sqrt{1-b^2} s^3 \notag \\
&&+(b-1) b s^4-(b-2)^2 (b-1) (b+1)^3 s^2+(b-1)^2 (b+2) (b+1)^6 \Big)\\
h_{71}(s,  b) &=& (b+1) \left(b^2 (-s)+2 b \left(\sqrt{1-b^2}-s\right)+2 \sqrt{1-b^2}+s\right)\\
h_{72}(s,  b,  \alpha) &=& 2 \alpha  (b+1) \left(b \sqrt{1-b^2}+\sqrt{1-b^2}+s\right)-h_{71}(s,  b)\\
h_{81}(s,  b) &=&(b+1)^4 \left(2 \sqrt{1-b^2} (b+1) s+(b-1) (b+1)^3-s^2\right)\\
h_{82}(s,  b) &=& (1-b) (b+1)^5 \left(2 \sqrt{1-b^2} (b+1) s+(b-1) (b+1)^3-s^2\right) f_1(s, b)^2
\end{eqnarray}
\begin{eqnarray}
h_{83}(s , b ) &=& \sqrt{1-b^2} s \Big(-\left(5 (b+2) b^2+b-2\right) f_1(s,  b) f_2(s,  b)-5 b (b \left(b^2+b-3\right)+1) \notag\\
&&(b+1)^3 f_2(s, b) +(b-1) (b+1)^4 f_1(s,  b)\Big)  -\sqrt{1-b^2} (b (b+2)-1)\notag\\
&& (b (b+3)+3) s^3 f_2(s, b)-2 (b+1)^2 \left(b^2 (b+2)-3\right) s^2 f_2(s, b)\notag\\
&&+(b-1) (b+1)^2 \Big((b-1) (b+1)^4 f_1(s,  b) -\big(b (b+10) f_1(s, b)+ f_1(s, b)\notag\\
&&+10 (b-1) b (b+1)^3 \big) f_2(s, b)\Big)\\
h_{84}(s,  b) &=& 4 \bigg((b+2) (b (2 b (b+2)+9)+6)  f_2(s, b)^2 s^4-2 (4 b+3) \sqrt{1-b^2} (b (b+3) \notag \\
&& +3)  f_1(s, b) f_2(s, b) s^3 +2 (b-1) (b+1)^2  \Big(2 (3 b+1) (b (b+3)+3) f_1(s, b)\notag \\
&&+(b+1) (8 b (b (b+2)+3)-3) f_2(s, b)\Big) f_2(s, b) s^2-\sqrt{1-b^2} \big((b (11 b+18)\notag \\
&&+6) f_1(s, b)+10 (b-1) b (4 b+3) (b+1)^3\big) f_1(s, b) f_2(s, b) s + (b-1) (b+1)^2\notag \\
&&\Big(35 (b -1) b^2 (b+1)^4 f_2(s, b)^2+20 (b-1) b (3 b+1) (b+1)^3 f_1(s, b) f_2(s, b)\notag \\
&& + \big(2 (b (13 b+10)+1) f_2(s, b)-(b-1) (b+1)^4\big)  f_1(s, b)^2 \Big) \bigg)\\
h_{85}(s,b,\alpha,\beta) &=&f_2(s,b) \bigg( \Big(5 b^5+10 b^4+b^2 \left(s^2-10\right) +b \left(3 s^2-5\right)+3 s^2\Big) f_2(s, b) \alpha \notag \\
&&-(b+1) f_1(s, b) f_2(s, b) \beta + \Big(6 b^4+8 b^3-4 b \big(\sqrt{1-b^2} s \notag \\
&&+2\big)-3 \sqrt{1-b^2} s-4 b^2-2\Big) f_1(s, b) \alpha \bigg)
\end{eqnarray}

\bibliographystyle{unsrt}

\end{document}